\documentclass[fleqn,twocolumn,10pt]{wlscirep}
\usepackage[utf8]{inputenc}
\usepackage[T1]{fontenc}

\usepackage{txfonts}
\usepackage{graphicx}
\usepackage{amsmath}
\usepackage{algorithm}
\usepackage{algpseudocode}
\usepackage{cuted}

\newcommand{\code}[1]{\texttt{#1}}

\DeclareMathOperator*{\argmin}{arg\,min}

\newcommand{\beginsupplement}{%
    \setcounter{table}{0}
    \renewcommand{\thetable}{S\arabic{table}}%
    \setcounter{figure}{0}
    \renewcommand{\thefigure}{S\arabic{figure}}%
 }

\newcommand{\topA}{
\overbrace{\hphantom{\begin{matrix} 
\cdots & \cdots & \cdots & \cdots &\cdots & \cdots  & \cdots 
\end{matrix}}}^{\mathcal{V}_{>=d}n_i}
}

\newcommand{\bottomC}{
 \underbrace{\hphantom{\begin{matrix} \cdots & \cdots & \cdots \end{matrix}}}_{(n-m-1)\cdot \mathcal{V}_{>=d}}}

\newcommand{\rightbrace}{
\left.\vphantom{\begin{matrix}0\\0\\0\\\ddots\\0\\0\end{matrix}}\right\}
\scriptsize\rotatebox{270}{$n_d-m$}}

\title{Learning Transport Processes with Machine Intelligence}

\author[*]{Francesco Miniati}
\author[ ]{Gianluca Gregori}
\affil[ ]{University of Oxford, Department of Physics, Parks Road, Oxford OX1 3PU, UK}

\affil[*]{francesco.miniati@pm.me}

\begin{abstract} {\bf [NEW]}
Transport processes ruled by complex micro-physics and
impractical to theoretical investigation may exhibit 
emergent behavior describable by mathematical expressions.
Such information, while implicitly contained in the results of
microscopic-scale numerical simulations close to first principles
or experiments is not in a form suitable  for macroscopic modelling.
Here we present a machine learning approach that leverages
such information to deploy micro-physics informed transport
flux representations applicable to a continuum mechanics 
description.
One issue with deep neural networks, arguably providing the most 
generic of such representations, is their noisiness which is shown to
break the performance of numerical schemes.
The matter is addressed and a methodology suitable for
schemes characterised by second order convergence rate is presented.
The capability of the methodology is demonstrated
through an idealized study of the long standing problem of 
heat flux suppression relevant to fusion and cosmic plasmas.
Symbolic representations, although potentially less generic,
are straightforward to use in numerical schemes
and theoretical analysis, and
can be even more accurate
as shown by the application to the same problem
of an advanced symbolic regression tool.
These results are a promising initial step to filling the gap
between micro and macro in this important area of modeling.
\end{abstract}

\begin{document}

\flushbottom
\maketitle
%
%
\thispagestyle{empty}


\section*{Introduction}\label{sec:outline}
    Conservation laws are fundamental laws of physics reflecting underlying symmetries of nature \cite{Noether1918}. In continuum mechanics they apply in a Lorentz invariant local form and are formulated mathematically as the continuity equation
    \begin{equation}\label{eq:cont}
        \frac{\partial u}{\partial t} = -\nabla\cdot \mathbf{q},
    \end{equation}
    where $u$, is the volume density of the conserved variable and 
    $\mathbf{q}(u)$ the corresponding current or flux density.
    Equation~(\ref{eq:cont}) states that
    the rate of change of a conserved variable within a volume $V$ is 
    due to its flux across the volume's surface, $\partial V$:
    \begin{equation}\label{eq:reynolds}
       \frac{d}{dt} \int_V u\, dV =  \int_{\partial V}  \mathbf{q} \cdot d\mathbf{s}.
    \end{equation}

    The continuity equation, typically in the form of a system of coupled partial differential equations (PDEs) for a state vector of conserved variables, $u$, 
    appears virtually in all fields of modern science and engineering where it is employed particularly for the 
    description of fundamental phenomena related to fluids, plasmas, and solids.
    The ability to obtain high fidelity models based on its accurate solution is therefore of great interest.
    Given its nonlinear character, the use of advanced numerical integration techniques for hyperbolic systems is usually required, which has given rise to a now well established and mature field of applied mathematics~\cite{Leveque2002,Allaire2007,cottet_koumoutsakos_2000}.

    However, the accurate knowledge of the transport term, $\mathbf{q}$, entering 
    the continuity equation is
    generally missing, particularly for diffusive processes 
    which depend on complex physics mechanisms operating at microscopic scales.
    This applies to many fundamental and industrial applications, including 
    fusion \cite{bell1981,Gregori2004,Brantov2013,Hu2016,McKelvey2017} 
    and cosmic plasmas \cite{Scudder2019,Komarov2018,Meinecke2021},
    liquid metals \cite{Viellefosse1975,Bernu1977},
    hypersonic flows during spacecraft re-entry \cite{Park2021},
    semiconductor devices \cite{Ichimaru1962,Jezouin2017}, {\bf  
    phononic transport in solids \cite{Phillpot2005,Qian2021}}.
    One outstanding example is the case of heat transport. For classical ideal fluids and gases it is well established that the heat flux is proportional to the temperature gradient as collisions between nearby particles enforce a local energy flow from hotter to colder regions. Thus Spitzer-H\"arm theory \cite{Spitzer1953} gives Fick's law, $\mathbf{q}=-\kappa\nabla T$, where $\kappa$ is the coefficient of thermal conduction and $T$ the temperature. 
    However, it has long been realised that the ideal gas approximation breaks down when the electron mean free path approaches or exceeds the temperature gradient scale-length, $L_T$, a condition common in thermonuclear fusion plasmas \cite{bell1981,Brantov2013}.
    Modified versions of Fick's law have been proposed in the literature but
    are often very poor and fail to generalize \cite{Gregori2004,Brantov2013}. 

    While such processes can in principle be modeled by integrating sets of 
    microscopic (e.g. kinetic) equations progressively closer to first 
    principles (which are, however, impractical to model macroscopic systems),
    this capability has unfortunately not yet translated into the formulation of
    transport terms, $\mathbf{q}$, employable in a continuum mechanics description,
    suitable for modeling macroscopic systems.
    In addition, there are physical conditions under which even current
    kinetic {\bf or {\it ab initio}} codes do not provide consistent results \cite{Grabowski2020}. 
    In this case it would be desirable to have the ability to learn about 
    such transport terms directly from actual experimental data.
    
    In this paper we describe a machine learning (ML) based approach designed 
    to improve our modeling capability and theoretical understanding of 
    generic transport processes by learning directly from data
    provided either by microscopic-scale numerical simulations or even experiments.
    In particular, we apply deep learning techniques to obtain a representation of the 
    transport process as a function of the state vector.
    
    In the past decade artificial intelligence has emerged as a powerful 
    technology \cite{LeCunn2015} and there has been great interests in its 
    use for scientific applications in general 
    \cite{Gamahara_2017,Dornheim2019,Erichson2019,Kasim2022,Pfau2020,Kasim2021}. 
    In the context of computational fluid dynamics machine learning has been leveraged as 
    an accelerator, {\it i.e.}, in order to enhance the performance of numerical solvers. 
    In particular, we have seen the development of powerful emulators, i.e. machines 
    capable of fully representing PDE solvers in order to reproduce the results of 
    conventional numerical simulation codes but at a significantly lower computational cost
    and/or higher accuracy \cite{Sanchez-Gonzales2021,wang2020,Zongyi2020,Rackauckas2020, Kim_2019,Lusch2018,Sanchez-Gonzalez2018}.
    Alternatively, researchers have focused on augmenting the modeling 
    capability of numerical methods. Here one typically employs a learnable 
    function to assist or replace altogether modular components of the 
    numerical scheme, particularly those most affected by finite resolution 
    effects, so as to enhance the overall performance of the method 
    \cite{Kochkov2021,Novati2021,pathak2020,SIRIGNANO2020,Um2020,Xie2019}.
    {\bf In the context of phononic thermal transport ML techniques have been applied
    to generate accurate effective (force field) potentials from high-fidelity density functional theory simulations \cite{Mueller2020,Qian2021}. In the spirit of the augmented methods discussed above, these effective potentials are then used in {\it ab initio} molecular dynamics simulations to predict heat conduction in new materials \cite{Roekeghem2016,Juneja2019,Qian2021}.}
    Note that pure ML emulators are not formulated on the basis of numerical 
    analysis. While as a result these solvers may be more flexible and 
    powerful as they are not subject to mathematical constrains as numerical 
    methods for hyperbolic systems (e.g., the Courant–Friedrichs–Lewy 
    condition), they usually come short of the stability, generalisation and 
    robustness characterising full fledged numerical methods 
    \cite{Leveque2002,Allaire2007,cottet_koumoutsakos_2000}. 
    These properties tend to be better preserved in the augmented methods.


    Our aim here is aligned with the development of augmented methods in that we 
    employ deep learning techniques to ultimately improve the accuracy of numerical 
    simulation models.
    At the same time our scope is different in that our 
    target is not a representation of the optimal numerical scheme, rather a 
    representation of the unknown underlying transport physics, {\bf somewaht 
    similar to the above phononic application}.
    Proper modeling of the latter usually requires a microscopic description based on
    a different (and much more expensive) computational approach or, at times, even 
    experiments.
    Our method therefore can also be seen as an accelerator in that it deploys a 
    micro-physics informed transport term usable in a conventional fluid approach 
    without incurring the cost of a full microscopic description.
    {\bf As will be shown in the next section, however, 
    a latent representation of a transport process provided by a deep neural network
    has limited smoothness properties, which breaks the performance 
    of a numerical scheme.
    For second order accurate schemes the issue can be addressed 
    by using latent representations of the flux gradient instead of 
    the flux function itself. This approach can be extended 
    for higher order schemes but at the cost of higher complexity.
    In any case, data regularization becomes now necessary
    to ensure a reliable estimate of 
    the gradients from the generally noisy data.}

    On an even more ambitious scale, one
    can attempt to obtain symbolic representation 
    of the transport process through a 
    symbolic regression analysis of the data.
    A mathematical expression is of great theoretical importance
    and allows for the potential discovery of new physical relational laws\cite{Cranmer2020,Udrescu2020}.
    In addition, it is straightforward to use in a numerical scheme. 
    So would it be the ultimate solution to the problem.
    Unfortunately, however, in general this task is significantly more difficult 
    and potentially strongly affected by the quality of the data\cite{Udrescu2020},
    
    In the remainder of this paper we first present
    in detail our proposed method 
    to learn representations of transport processes, 
    {\bf addressing the numerical issues, the 
    the data regularization procedure
    and the employed deep learning methods.
    In the second part we demonstrate the viability of the method in a scenario of real
    scientific interest, the case of heat transport in a high temperature plasma previously discussed, including the application of an advanced symbolic regression tool.}

\section*{Methodology}
    \subsection*{Preliminaries}
    Building a ML model of a transport process requires
    a set of data representing the solution to the associated equation~(\ref{eq:cont}).
    The data can be obtained either through direct 
    measurements (experiments), numerical simulations with sufficient 
    modeling capability, or both. In the former case the data most 
    likely represents time variations of the conserved quantity 
    and in the
    latter case values of the flux density itself.
    Obviously the data should only contain information
    about the transport process of interest.
    For example, in studying diffusion the effects of advection must be subtracted out.
    More precisely, our data is laid out on a discrete grid, $\Gamma \in \mathbb{Z}$,
    defined in a one-dimensional spacial domain $[0,L]\in \mathbb{R}$.
    This simple setting is not restrictive in terms of information scope 
    and conveniently keeps the data complexity to a minimum.
    Given the mesh spacing, $h$, we define a set of control 
    volumes $i \in \Gamma$ corresponding to region of space $[i\,h,(i+1)\,h]$ with boundaries 
    belonging to a face-centered discretisation space based on those control
    volumes: $\{\Gamma^{e} = i + 1/2 : i \in \Gamma \}$.
    It is convenient to relate the time variation of the conserved quantity
    and the fluxes
    by integrating equation~(\ref{eq:cont}) in a space-time slab ($h, \delta t$)
    \cite{godunov59}
    \begin{equation}\label{eq:div}
    \bar q_{i+\frac{1}{2}}-\bar q_{i-\frac{1}{2}} = - \frac{h}{\delta t} \delta\bar u_i,
    \end{equation}
    where $\bar q$ is the time averaged flux over $\delta t$, and $\delta 
    \bar u_i$ is the space averaged time variation of $u$ inside the $i$-th 
    control volume during $\delta t$.
    Note that equation~(\ref{eq:div}) at this stage is exact.
    Then knowing the value of the flux $q$ at one of the boundaries, 
    $b=0,L$, one can write:
    \begin{equation}\label{eq:flx}
    \bar q_{k+\frac{1}{2}} = q_b - \frac{h}{\delta t}\sum_{i=m, n} \delta \bar u_i
    \end{equation}
    where $(m, n, b)=(0,k, 0)$ or $(k,L,L)$ if the flux value is known at 
    the domain's origin or end, respectively.
    Equation~(\ref{eq:flx}) is useful when the value of the flux, $\bar q_{i+1/2}$'s, is not 
    directly measurable, in which case it can still be inferred from the 
    $\delta \bar u_i$'s.
    It also makes it clear that the fluxes and time variations 
    we deal with are actually time and volume averages not instantaneous or 
    point-wise values. Similarly, although we use a one-dimensional model, 
    experimental data will require a surface averaging of the flux. 
    Since in general such experimental and numerical inaccuracies can be 
    quantified there is in principle control over the quality of the data
    and the inferred model. 
    With this understanding the data, either the set of $\delta \bar u_i$'s or equivalently of 
    $\bar q_{i+1/2}$'s, can be used to construct the
    labels for supervised training. We 
    address details related to this step next.

    \subsection*{Smoothness {\bf [NEW]}}
    The performance of a numerical scheme
    depends on the smoothness of the functions
    appearing in the target PDEs.
    While this is not an issue with analytic expressions unless the functions
    of interest are intrinsically irregular, the question arises in the case of
    latent representations provided by a Multilayer-Perceptron (MLP),
    owing to their noisy character.
    The smoothness we are referring to here concerns
    the validity of Taylor expansion's approximation, from which  
    the accuracy and convergence properties of a scheme are
    inferred through the methods of numerical analysis\cite{Allaire2007}. 
    We thus assess the smoothness of an MLP function by the
    residual error, $\mathcal{E}_p$, 
    of its Taylor expansion up to order $p-1$, 
    for variation $h$ of an individual components $\xi$ of ${\bf x}$,
    namely
    \begin{eqnarray} \label{terr.eq}
    \mathcal{E}_p (h; {\bf x}_0, f)=\frac{1}{f_0}
    \left(f({\bf x}_0+h{\bf \hat \xi})-\sum_{k=0}^{p-1} 
    \frac{\partial^k_\xi f_0}{k!} h^k \right) = O(h^p)
    \end{eqnarray}
    where ${\bf x}_0$ is the expansion point, $\partial^k_\xi f_0\equiv
    \partial^k_\xi f({\bf x}_0)$
    and ${\bf \hat \xi}$ is a unit vector in the $\xi$-direction of parameter space.
    The last equality is expected to hold for any $p$ for 
    infinitely differentiable functions such as the ones 
    we are dealing with.
    We compute $\mathcal{E}_p$ 
    for the analytic function $q({\bf x}=(n,~T,~\beta))$ defined in 
    equation (\ref{eq:qkom}),
    and for its MLP representation described in the 
    `MLP Representation of the Flux Function' in Methods, 
    although the specific details do not affect the conclusions of
    the current discussion.
    
\begin{figure}[ht]
        \centering
        \includegraphics[width=1.0\textwidth,]{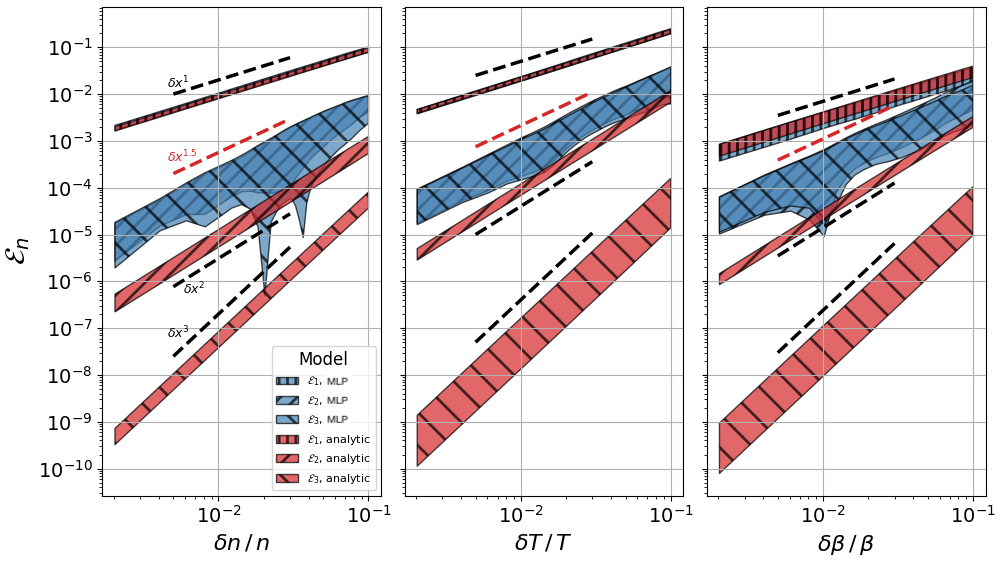}        
        \caption{
        Taylor expansion's error:
        residual error from Taylor expansion up to order 
        0, 1, 2 (corresponding to $\mathcal{E}_p$ with $p=1, 2, 3$, 
        respectively in the legend) of the analytic flux function 
        $q$ in equation~(\ref{eq:qkom}) (red) and its MLP representation (blue).
        Each panel corresponds to 
        variations of each individual thermodynamic variable,
        with the shaded regions representing
        the range between the absolute of the mean value
        (bottom boundary) and one standard deviation (upper boundary)
        for a sample of 30 randomly chosen expansion points.
        \label{Fig1}}
    \end{figure}

    In Fig.~\ref{Fig1} we plot
    $\mathcal{E}_p(h)$ as a function of $\delta \xi/\xi\equiv h/x_{0,\xi}$,
    after averaging over 30 expansion points, ${\bf x}_0$, randomly chosen
    in the domain of the $q$ function for $p=1,~2,~3$.
    Each panel corresponds to the expansion along a different
    component $\xi$, with the shaded regions representing
    the range between the absolute mean value
    (bottom boundary) and one standard deviation (upper boundary).
    The results indicate that while for the analytic case (red),
    $\mathcal{E}_p\propto h^p$, as expected, 
    for the  MLP representation 
    (blue) only $\mathcal{E}_1$ follows expectations, in fact
    overlapping with the analytic counterpart.
    Instead, $\mathcal{E}_2$ 
    scales only as $h^{1.5}$
    and $\mathcal{E}_3$ is virtually equivalent to $\mathcal{E}_2$ 
    (notice the cross hatch pattern),
    suggesting that additional terms of Taylor's series do not
    improve the approximation.
    
\begin{figure}[ht]
        \centering
        \includegraphics[width=1.0\textwidth,]{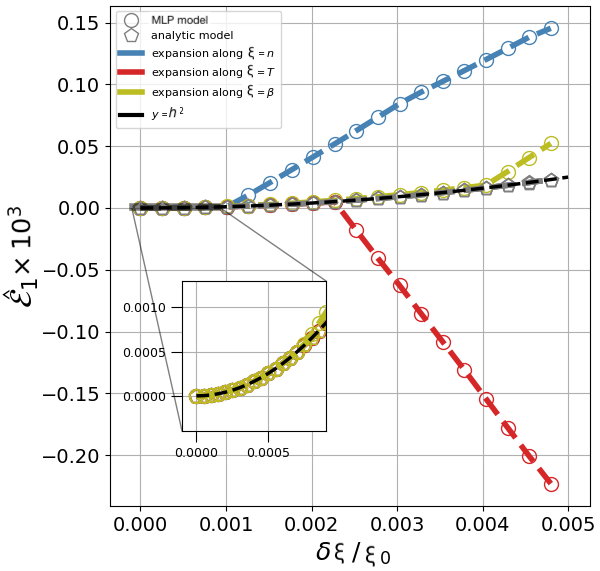}        
        \caption{Smoothness test:
        plot of the scaled residual of a first order Taylor expansion
        $\mathcal{\hat E}_1= \mathcal{E}_2 \,/\,\mathcal{E}_2(h\approx 10^{-3}x_{0,\xi})$,
        with respect to each thermodynamic variable, $n, T, \beta$ (no sample averaging over
        the expansion point was computed).
        The open gray pentagons represent the three overlapping expansions 
        for the analytic flux function $q$ in equation~(\ref{eq:qkom}).
        The blue, red and olive symbols (circles + dash-line) 
        show the expansions with respect to 
        $n, T, \beta$, respectively, for the case of the MLP representation.
        The black solid line is the parabolic curve that all scaled residual errors 
        are expected to follow, an expectation fulfilled by the analytic case,
        but not the MLP representation, except that for a very short interval.
        \label{Fig2}}
    \end{figure}

    In Fig.~\ref{Fig2} we plot $\mathcal{\hat E}_2$, 
    i.e. $\mathcal{E}_2$
    computed for a single value ${\bf x}_0$ but rescaled by the factor
    $\mathcal{E}_2(h\approx 10^{-3}\,x_{0,\xi})$.
    The figure shows
    that the scaled Taylor expansion of the MLP representation 
    does follow the parabolic curve (black line), 
    but only within a much smaller interval than the analytic case (gray pentagon).
    In the specific case, beyond $\delta \xi/\xi \leq 10^{-3}$, 
    $\mathcal{\hat E}_2$ grows linearly indicating that the 
    function's derivative has changed substantially.
    That this simple description, indicative of a noisy character,
    is sufficient to reproduce the qualitative and quantitative
    behaviour of the MLP representation illustrated
    in Supplementary Fig.~\ref{FigA1}. Here
    the same sample averages of $\mathcal{E}_p$ as in Fig.~\ref{Fig1}
    are plotted together with those of the analytic function
    $q$ modified for an additional random term proportional to its first derivative,
    namely
    \begin{equation} \label{eq:stochgrad}
    q_s({\bf x}) = q({\bf x}) + a\, \nabla q \, \frac{h^{1.5}}{x_{0,\xi}^{0.5}},
    \end{equation}
    where $a$ is a random number
    sampled uniformly within $[-A, ~A]$ with $A$ of order unity.

    This lack of smoothness implies poor numerical performance.
    For example, applying a finite difference centered scheme
    to estimate the gradient of $q$ yields only $O(h^{0.5})$
    accuracy as opposed to $O(h^2)$ as usual.
    The resulting truncation error is likewise of order $\tau(h)\propto O(h^{0.5})$.
    Oddly enough, however, due to the stochastic character of the
    offending term spoiling Taylor expansion's approximation
    the convergence rate would be better than inferred by the truncation
    error because the error accumulates only as $N_{steps}^{1/2}$. 
    Therefore, in a finite different scheme for example, 
    at a given solution time, $t=N_{\rm steps}\;\Delta t$ and for fixed
    $\Delta x /\Delta t =$const.
    \begin{equation} \label{eq:erranalysis}
    \varepsilon(t,\Delta x, \Delta t)= 
    \sum_{N_{\rm steps}} \tau(\Delta x)\; \Delta t
    \propto \tau(\Delta x)\; (\Delta t)^{1/2} = O(\Delta x).
    \end{equation}
    Later in the paper we will support this finding with an actual
    numerical experiment.
    
    First order convergence rate would still be poor by modern standards.
    However, in view of the foregoing discussion 
    it is clear that using an MLP representation of the flux gradient 
    instead of the flux itself would suffice for the purpose of 
    a second order accurate scheme.
    In fact, the primitive of the MLP representation, the flux function,
    would now have a valid Taylor expansion up to $p=2.5$ and, by the above arguments,
    would be able to fulfill the requirements of second order convergence rate.
    Suitably smooth functions for higher order schemes could be also built
    starting from representations of correspondingly higher derivatives,
    although it would require additional integration operations further
    complicating the overall scheme. Thus, in the `Results' section we 
    stick to second order accurate schemes.

    \subsection*{Regularisation {\bf [NEW]}}
    The presence of noise in the data prevents direct calculation of
    the flux gradient (and higher derivatives).
    To improve the data quality we therefore first undertake a
    regularization procedure. Tikhonov's method was found particularly 
    effective for this purpose\cite{Tikhonov1963,Cullum1971,Eilers2003,Chartrand2005,Knowles2014}. 
    The method solves an optimization problem in which,
    given a set of noisy data $y_i({\bf x}_i)$, 
    the smoothed, noise reduced data, $\hat y$, is found by minimising an objective function, $Q$, containing two contrasting terms, one 
    measuring $\hat y$'s fidelity to the original data and the other 
    its smoothness.
    We evaluate the former term through the mean-squared-error (MSE)
    and the second from its high order 
    derivatives ,$y^{(p)}$, 
    as estimated from the regularised values, i.e.
    \begin{equation} \label{eq:qsim}
    Q((\hat{y}, y)  
    =  \|y - \hat{y}\|^2 + \lambda_s \sum_p \|\rm \hat{y}^{(p)}\|^2,
    \end{equation}
    where $\lambda_s$ is a parameter weighting the regularization term.
    Although equation (\ref{eq:qsim}) is typically formulated in the context
    of one-dimensional data, we apply it volumetrically,
    i.e. simultaneously to all different ${\bf x}$-space components.
    Thus ${\bf x}, y$ and $y^{(p)}$ correspond {\it linearised} 
    one-dimensional data array.
    While increasing the computational complexity, 
    this yields isotropic smoothness of $y({\bf x})$.
    Using matrix operators
    \begin{equation} \label{eq:qgen}
    Q((\hat{y}, y) =
    (\rm \hat{y}-y)^{\rm T}U(\hat{y}-y) + \lambda_s 
    (\mathcal{D}^p\hat{y})^{\rm T}U(\mathcal{D}^p\hat{y}).
    \end{equation}
    where U weights the element-wise contribution 
    according to its volume measure in the MSE metric
    (i.e. the $d{\bf x}$ element in an integral)
    and $\mathcal{D}^p$ consists 
    of a stack of partial differential operators with
    respect to all components of ${\bf x}$ including, in general, 
    mixed terms up to order $p$ (see 'Regularisation' in Methods for additional details).
    In the application example discussed in the next Section, where
    $y$ is the heat flux and ${\bf x}$ the thermodynamic variables,
    we find it sufficient to include only non-mixed terms of order $p=4$.
    The optimal set, $\hat y$, is found
    by solving the equation obtained by 
    setting $\partial Q/\partial \hat y$ to zero, hence
    \begin{equation} \label{eq:yhat}
    \hat y = \rm (I + \lambda_s 
    \mathcal{D}^{p^{\rm T}} U \mathcal{D}^p)^{-1} y.
    \end{equation}
    As for the $\lambda_s$ parameter, we follow the practice of 
    setting its value such that the resulting $(y - \hat{y})$ 
    statistics is close to what is expected for the random 
    errors of the data. In particular, when the noise standard
    deviation, $\sigma_y$, is known, Morozov\cite{Knowles2014} principle can be 
    applied and 
    \begin{equation}
    \lambda_s = \argmin_{\lambda_s} \left(\sigma_{\hat y}(\lambda_s) -\sigma_y\right)^2.
    \end{equation}
    When $\sigma_y$ is unknown, generalised 
    cross-validation method\cite{Eilers2003} can 
    be applied instead. In our experiments we find these two methods
    to give virtually identical results and use
    Morozov's for computational convenience.

    \subsection*{Latent Representation {\bf [NEW]}}
\begin{figure}[ht]
        \centering
        \includegraphics[width=0.9\textwidth]{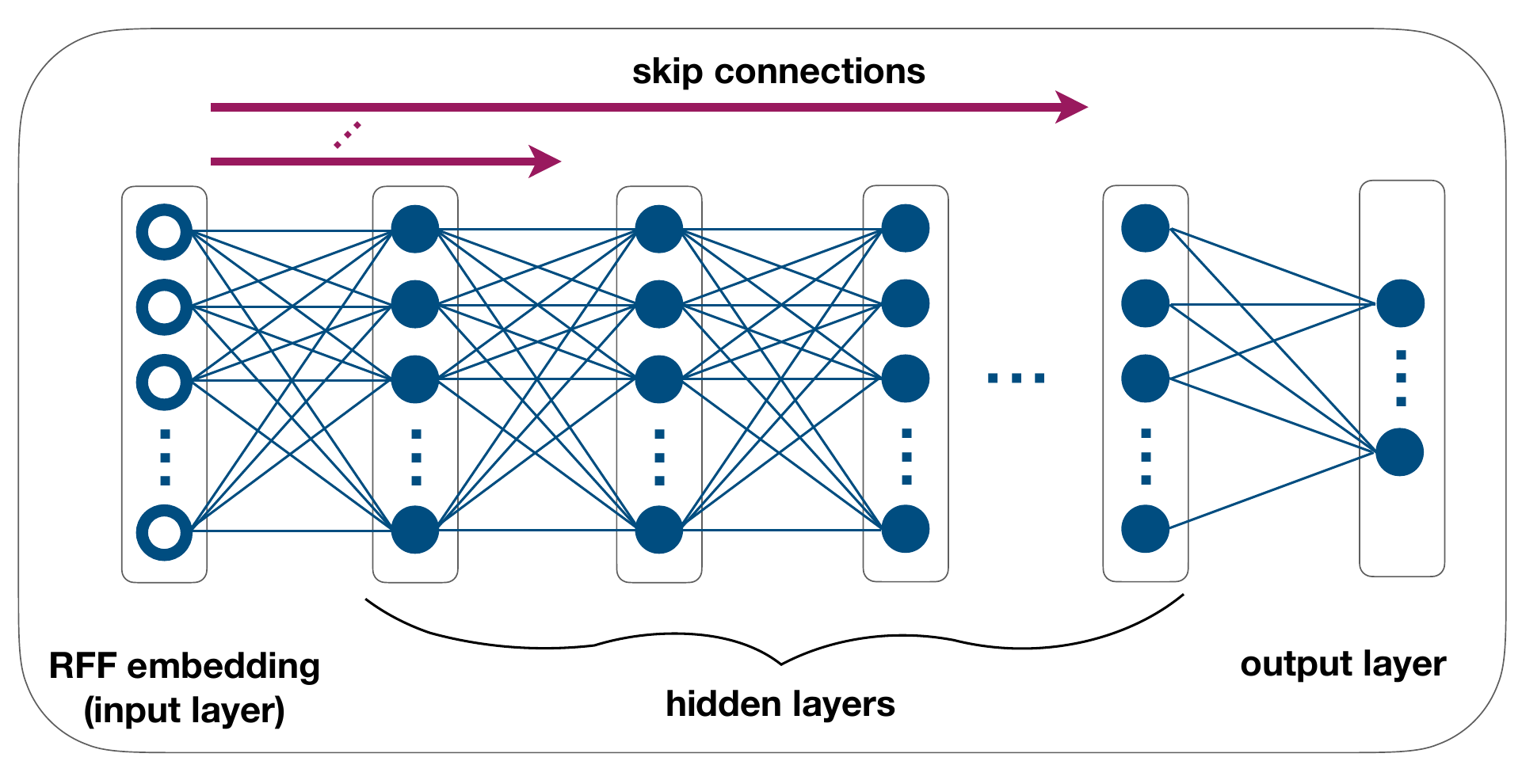}
        \caption{Trainable MLP representing the flux-gradient function. The first layer embeds the input features via RFFs which are then fed to the first hidden layer. The number of RFFs is the same as the number of hidden units which is constant across the hidden layers.
        Nonlinearity is introduced by application of a ReLU activation function to the affine mapping returned by the hidden units. The RFF embeddings are also fed to every other hidden layer except the last through skip connections. The output layer consists of as many
        regression units as the gradient components without activation function.
        \label{Fig3}
   }
    \end{figure}

    In order to obtain a latent representation of the 
    transfer process of interest we use an MLP 
    with the architecture summarised in Fig.~\ref{Fig3} and 
    described in detail in `MLP Architecture' in Methods.
    The MLP is trained to learn the gradient of the flux function 
    using a set of labels obtained by differentiating 
    the regularised input flux data, as described in the previous section.
    The training is based on Stochastic Gradient Descent using
    an objective function given by the MSE of the relative error
    of the predicted value with respect to its label 
    (further details are given in `MLP Architecture' in Methods).
    
    While in our application example discussed later on
    the flux density and its gradient depend on the point-wise
    values of the thermodynamic variables, a more general dependence 
    from values in a neighbouring region of the evaluation point 
    may at times be required. For such case an MLP may not suffice
    and a more flexible implementation of our machinery is conceivable, 
    in which our MLP unit is embedded within a non-local-neural-network 
    or even a full graph-network\cite{Battaglia2018,wang2020,Wu2021}
     (of which the former is a special type).

\section*{Application to Heat Transport} \label{sec:application}
    \subsection*{Basics}
    We now consider the case of 
    heat transport in a high temperature plasma, a case
    of realistic scientific and engineering interest.
    As already mentioned in the Introduction, classically this process is described by Fick's law with a thermal diffusion coefficient given by Spitzer-H\"arm model~\cite{Spitzer1953}
    \begin{align} \label{eq:qsh}
        q_{SH} & =-\kappa\, T_e/L_T, \quad
        \kappa = \frac{128}{3\pi} \zeta n_e v_{te} \lambda_{ei}, 
        \\ 
        v_{te}  &= \left(\frac{2T_e}{m_e}\right)^\frac{1}{2},  \; 
        \lambda_{ei} =\frac{3T_e^2}{\sqrt{32\pi}Z^2n_e e^4 \Lambda}, \; 
        \zeta =\frac{0.24+Z}{4.2+Z},
        \nonumber
    \end{align}
    where $T_e,\, v_{te},\, n_e,\, \lambda_{ei}$ are the electron 
    temperature, thermal speed, number density and collisional mean
    free path, respectively, $L_{T} \equiv T_e/\nabla T_e$ is the 
    temperature gradient scale-length, $Z$ is the ions charge state, $e$
    the electric charge, $m_e$ the electron mass, and $\Lambda$ the Coulomb logarithm. 
    The model equations \ref{eq:qsh}, 
    become invalid (and the heat flux strongly suppressed) as the electron mean free path 
    approaches the temperature gradient scale-length, i.e. 
    $\lambda_{ei}\nll L_{T}$\cite{bell1981,Gregori2004,Brantov2013}.
    Kinetic models based on a phase-space 
    description of the plasma continue to apply 
    so one possibility would be to learn the representation of 
    latent heat flux function from such simulation data.
    For simplicity, however, 
    in the following we use a dataset of values generated
    from the following heat flux function~\cite{Komarov2018}
    \begin{eqnarray} \label{eq:qkom}
        q(n_e,T_e,B)= n_eT_e v_\parallel = \epsilon\, n_eT_e v_{te}, \quad \epsilon(n_e,T_e,B)\equiv\frac{v_\parallel}{v_{te}},
    \end{eqnarray}
    where $B$ is a magnetic field strength, $v_\parallel$ is the velocity component parallel to the magnetic field and 
    \begin{eqnarray} \label{eq:eps}
        \epsilon(n_e,T_e,B) =
        \left(\frac{L_{T}}{\lambda_{ei}}+\beta_e+4\right)^{-1}, 
        \quad \beta_e=\frac{n_eT_e}{B^2/8\pi}. \label{eq:beta}
    \end{eqnarray}
    yielding a suppression factor $$\epsilon \,L_T/\lambda_{ei}$$ with
    respect to
    Spitzer-H\"arm's flux, $q_{SH}=n_eT_ev_{te}\lambda_{ei}/L_T$.
    
    Physically, this model describes the heat flux suppression due to 
    whistler instabilities occurring in a low density, high-$\beta$ 
    intergalactic plasma, characterised by $n_e\approx 10^{-4}$ cm$^{-3}$,
    $T_e\approx$ a few keV, and $B\approx 10^{-6}$ G, in the presence of 
    temperature gradients with scales $L_T\approx 10^{22}$ cm \cite{Komarov2018}.
    In other words the above equations describe a specific
    emergent behavior of the heat flux suppression mechanism,
    caused by specific complex microscopic processes operating over several $\lambda_{ei}$ scales.
    As already pointed out, however, the heat flux suppression is a
    phenomenon occurring whenever $\lambda_{ei} \nll L_T$
    irrespective of the underlying mechanism responsible for it.
    So for conditions relevant to, e.g., High Density Plasma 
    Physics and Inertial Confinement Fusion, with similar keV temperatures 
    and $n_e \approx 10^{19}-10^{23}$ cm$^{-3}$ but not necessarily
    supporting ordered magnetic fields,
    there would still be a transition to non-local transport for 
    a temperature scale $L_T\approx 10^{-1}-10^{-5}$cm 
    although the driving physical mechanism may differ 
    \cite{Brantov2013,Gregori2004}. 
    Likewise can be said of plasmas characterised by different 
    parameters that still combine to produce a similar value of $L_TT_e^2/n_e$.
    In each of these cases the cause and specific characteristics 
    of the emergent behaviour will be different.
    However, our purpose here is to demonstrate that if such an
    behavior exists and can be described in terms of a set of input parameters
    characteristic of the plasma state, then we shall be able to capture it.

    \subsection*{Datasets {\bf [NEW]}}

    To build the various datasets for the supervised training
    we consider a domain defined by the parameter ranges in
    Table~\ref{tab:plasmaparams}
    and discretise it with a uniform grid.
    At each grid point, ${\bf x}=(n_e,~ T_e,~ \beta_e)$,
    we then evaluate the flux function $y=q({\bf x})$ according to equation (\ref{eq:qkom}).
    To assess the impact of the data volume on the model's 
    performance we have
    generated three sets of data with different sampling density described
    by the number-of-points-per-decade parameter, $N_{ppd} = 5,~ 10,~ 20$.
    As detailed in the next Section the computed gradient function
    share the same grid as the flux function
    except for a one-point surface layer
    where the gradient cannot be fully computed.
    To compensate for this we pad the grid surface
    with a 1-point buffer zone. This surface buffer zone is then doubled
    to also account for an additional layer of gradient data that
    we do not use due to the degrading performance of the 
    regularization step there.
    Thus, in general the grid dimensions are,
    $(N_n, N_T, N_\beta)=(2N_{ppd}+4, ~N_{ppd}+4, ~2N_{ppd}+4)$, 
    with a corresponding grid spacing 
    $\Delta \xi=(\xi_{max}-\xi_{min})/(N_{\xi}-1)$, with the min and max values
    given in the Full Grid section of Table~\ref{tab:plasmaparams},
    and $N_\xi$ the grid size corresponding the parameter $\xi$.
    Notice that because of the different grid spacings
    for different values of $N_{ppd}$
    the actual parameter range covered by the gradient data
    differs for the different datasets, as illustrated by the 
    `Gradient Grid PR-$N_{ppd}$'
    sections in the lower part of Table \ref{tab:plasmaparams}.
    The flux suppression factor $\epsilon L_T/\lambda_{ei}$,
    also computed in the table, continues nonetheless
    to range from values $\ll 1$, i.e. the regime of highly suppressed flux,
    to values $\approx 1$, corresponding to the Spitzer-H\"arm limit.
    Finally, to assess the impact of noise in the data, we also
    generate datasets in which the flux-density values are modified to include
    a normally distributed random percentage error,
    \begin{eqnarray}
        y &\gets& (1+\hat \sigma_n)  y
    \end{eqnarray}
    with $\hat \sigma_n$ a random variate from a Gaussian distribution $\mathcal{N}(0, \sigma_n)$.
    
    The list of datasets is summarised in Table \ref{tab:data}. The first column is the 
    name of the dataset and the second the value of $\sigma_n$ multiplied by 100. The next three columns indicate the $N_{ppd}$ parameter, the buffer size, 
    and the total number of flux function evaluations, respectively.
    The last four columns refer to the flux gradient, 
    in particular the total number of evaluations and the size of the 
    training, validation and testsets partitions, respectively.

    \subsubsection*{Data Regularization {\bf [NEW]}}
    For each dataset in Table \ref{tab:data} 
    we apply Tikhonov's regularization to the log values of $y$ as a 
    function of the log values of $x$. This means that our fidelity term
    in equation (\ref{eq:qsim}) is a relative error. The regularization term 
    is given by the 4-th derivative computed on the regularised data
    based on finite differences of adjacent cell values 
    (we effectively apply the operator $\cal{D}^{4,1,*}_{N_n, N_T, N_\beta}$ 
    described in `Regularization' in Methods).
    The flux gradient, providing the labels for 
    the MLP's supervised learning discussed above,
    is then computed using a second-order accurate central difference scheme
    (the operator $\cal{D}^{1,2,*}_{N_n, N_T, N_\beta}$)
    on the regularised data. Thus, for the $\xi$ component
    \begin{equation}
    y^{(1)}_\xi({\bf x}_i) = 
    \frac{\hat y({\bf x}_i+\Delta \xi\, {\bf \hat \xi}) - \hat y({\bf x}_i-\Delta \xi\,{\bf \hat \xi})}{2\Delta \xi}.
    \end{equation}
    
\begin{figure}[ht]
        \centering
        \includegraphics[width=1.0\textwidth,]{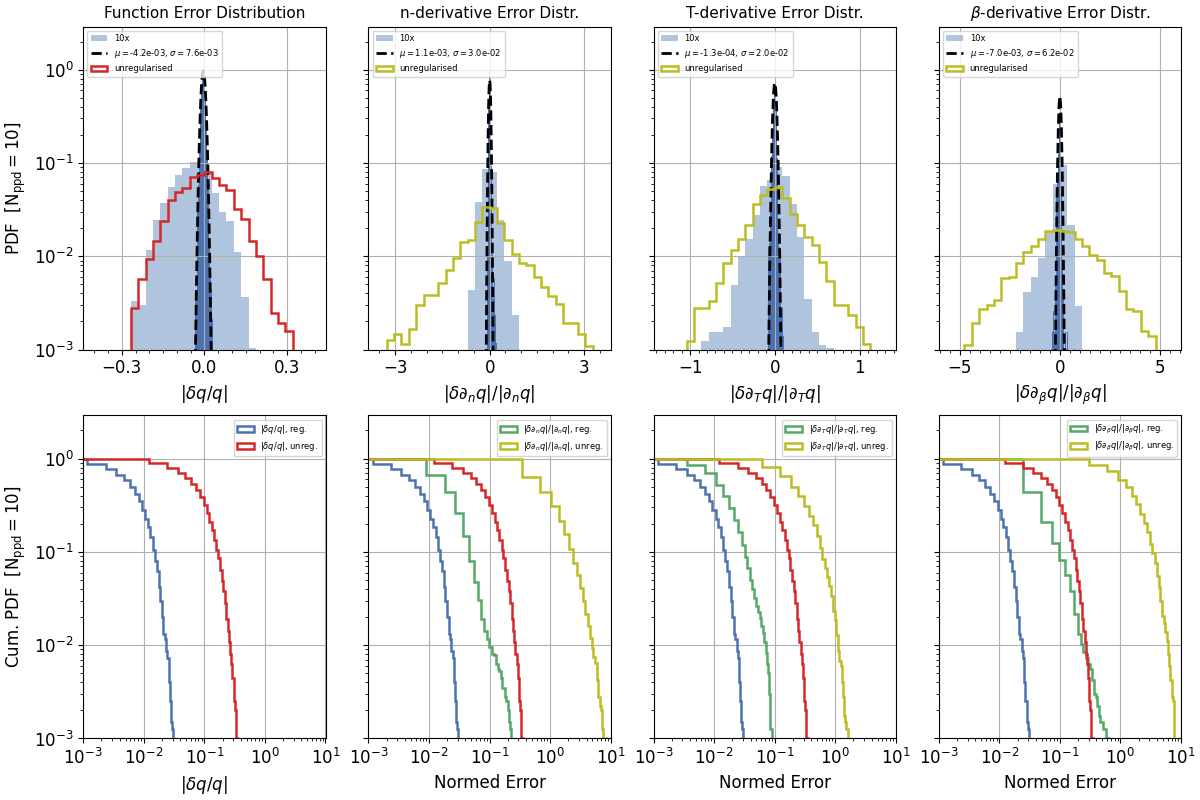}        
        \caption{Single regularization result:
        Regularization results for the dataset B.10, with $N_{ppd}=10$ and $\sigma_n=0.1$.
        Top panels: histograms of the 
        relative error distribution of the regularised and unregularised  
        flux function (left) and each gradient component (next three panels), 
        respectively (see legend for details). The blue shaded regions correspond
        to the regularised error distribution expanded by a factor 10.
        Bottom panels: corresponding cumulative error distributions for the 
        histograms in the top panels, particularly 
        the regularised and unregularised flux function data (blue and red)
        and its individual gradient components (green and olive), respectively.
        \label{Fig4}}
    \end{figure}

    Fig.~\ref{Fig4} shows the results of the regularisation procedure
    in terms of the accuracy of the flux function and its gradient component, 
    for the specific case of the dataset B.10,
    i.e. $N_{ppd}=10$ and $\sigma_n=0.1$.
    The top panels from left to right show the relative error distribution of the 
    regularised and unregularised
    flux and of each of its gradient component, respectively.
    The blue shade corresponds to the regularised error
    distribution expanded by a factor 10.
    The bottom part similarly compares, in the corresponding panels,
    the cumulative error distributions
    of the regularised and unregularised flux function (blue 
    and red) and its gradient components (green and olive), respectively.
    The plot shows that Tikhonov's regularization is very effective at 
    suppressing the noise in the original data allowing reliable estimates
    of the function gradients even in the case of relatively high noise. In this 
    particular case we effectively obtain errors RMS below 1\% and of order
    of a few \% for the flux function and its gradient components, respectively,
    starting from a dataset with 10\% normally distributed relative error.
    The improvement is particularly dramatic for the  
    flux gradient whose calculation, as is well known, would be otherwise
    very challenging.
    
\begin{figure}[ht]
        \centering
        \includegraphics[width=1.0\textwidth,]{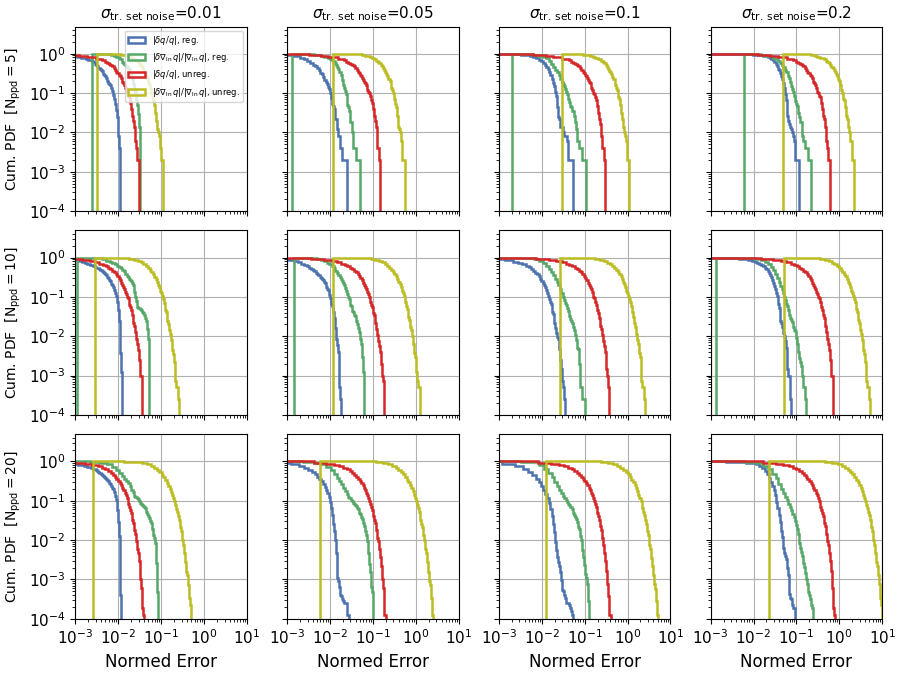}        
        \caption{Regularization results:
        for each combination of the $N_{ppd}$ and $\sigma_n$ parameters,
        the blue and red curve show the cumulative distributions of the relative error
        of the flux function while the green and olive curves show
        the cumulative distributions of the flux gradient relative error, 
        estimated by the ratio of the Euclidean norm of the flux log-gradient error
    and the Euclidean norm of the correct flux log-gradient (see main text
    for definition of log-gradient and Euclidean norm).
        \label{Fig5}}
    \end{figure}

    The results for all datasets are summarised in Fig.~\ref{Fig5}.
    For each combination of the $N_{ppd}$ and $\sigma_n$ parameters,
    the various panels show the cumulative distribution of the relative error 
    of the flux function and its gradient, for the the regularised (blue and green) 
    and unregularised (red and olive) data, respectively.
    To estimate the flux gradient error, 
    we compute the Euclidean norm of the flux log-gradient error and
    divide it by the Euclidean norm of the correct flux log-gradient.
    Here log-gradient of $f({\bf x})$, with ${\bf x}$ a vector,
    means that $(\nabla_{\log {\bf x}}f)_i=\nabla_{\log x_i}f$ and
    Euclidean norm of ${\bf x}$ means $(\sum_i x_{i}^2)^{1/2}$.
    Additional statistics on the errors and their trends with the 
    $N_{ppd}$ parameter are shown in Supplementary Fig.~\ref{FigA2}.
    Consistently with Fig.~\ref{Fig4}, Fig. \ref{Fig5} and Supplementary Fig. \ref{FigA2}
    show the effectiveness of the
    regularization, which allows us to compute estimates of the flux gradient
    with an accuracy that in most cases is significantly better than even 
    that of the original data.
    It is tempting but somewhat not straightforward to compare the panels corresponding to the 
    same $N_{ppd}$. In fact, on the one hand as already pointed out the domain
    of the regularised data differs for different values of the $N_{ppd}$ parameter.
    On the other hand, the higher $N_{ppd}$, i.e. the grid resolution, 
    the higher the impact of noise on the calculation of the gradient components,
    as is clearly visible in Supplementary Fig. \ref{FigA2}.
    In the following we will show that it is definitely
    advantageous to use finer grids, i.e., larger $N_{ppd}$.
    
    \subsubsection*{Training Data}
    The MLP is trained on a set of data consisting of features, ${\bf x}$, the
    thermodynamic variables already discussed,
    and labels ${\bf y}^{(1)}$ corresponding to the gradient of the flux function
    obtained from the regularised data.
    We take the natural logarithm of both except for the $\beta$
    component of the flux gradient for which we take the log of the negative
    value, and normalise both ${\bf x}$ and the ${\bf y}^{(1)}$ components 
    to have zero mean and unit standard deviation.

    \subsection*{Learned Representations {\bf [NEW]}}

    For each dataset listed in Table~\ref{tab:data} we train 
    a total of 100 MLP-models with hyperparameters selected from 
    a (reduced) search space given in Table~\ref{tab:hpsearch} 
    (see `Hyperparameter Optimization' in Methods for further details).
    Table~\ref{tab:models} shows a selection of best models with 
    the corresponding used dataset, hyperparameters and also final RMS and Max 
    {\it evaluation} errors. 
    We aimed for evaluation errors to be at least 
    consistent with those characterising the regularised data.
    We have repeated the analysis and tests shown in this section with 
    alternative selection of best models obtained during the hyperparameter search 
    and found consistent results.
    In the following we compute various statistics of the model prediction errors
    in which all gradient components are treated without distinction.
    This is feasible because the models are trained to learn the log of the labels, 
    and the prediction errors correspond to relative errors (hence the axis label).
    It also means that the computed error refer equally to each component.

    Fig.~\ref{Fig6} shows a summary of the {\it test-errors} 
    for the models in Table~\ref{tab:models}. 
    For each model of the A-, B-, C- series the errors are computed with
    respect to the noiseless testset, i.e. the testsets of the A.0, B.0 and C.0
    datasets respectively.
    Each row corresponds to a different $N_{ppd}$ (and 
    parameter range domain PR-$N_{ppd}$), while each column corresponds to a 
    different value of $\sigma_n$, the noise in the flux dataset before applying
    Tikhonov's regularization.
    The bias $\mu$ is typically negligible with respect to the variance term and,
    as it would be expected, in general the performance improves consistently
    for smaller values of the noise, $\sigma_n$, and for denser datasets, i.e. larger $N_{ppd}$.
    It is interesting to note that for large values of the input noise, 
    i.e. $\sigma_n =10, 20$, the
    MLP representations are characterised by an RMS error per component 
    $\simeq \sigma_n /2
    \simeq 2 \sigma_{reg}$, i.e. half the pre-regularization
    error but twice as large that of the regularised data (Supplementary Fig. \ref{FigA2}).
    This is likely a consequence of the fact that the residual error noise in the
    regularised data is not purely Gaussian and the MSE 
    is a less effective objective for closing in on the ground truth.
    Here, we also see that although in Supplementary Fig.~\ref{FigA2}
    the error values for the regularised data at given percentile appear to plateau,
    MLP representations trained with more densely sampled datasets always 
    shows an overall better performance. 
    Below we show a more direct comparison of the models
    tested within the same parameter domain.
    In any case it is remarkable that we can obtain representations of the flux
    gradient with errors per components of only a few percent, despite an input
    relative error of the flux data of even 10\%.
    
    Fig.~\ref{Fig7} shows more specific error statistics,
    in particular the RMS (blue dash line), 
    Max (red dash line) and Bias values (yellow thin dash line),
    characterising the model predictions and their trend with the
    pre-regularization noise.
    From top to bottom, the rows correspond to errors computed using 
    the noiseless testset of the C-, B- and A- Series respectively,
    while from left to right the columns
    correspond to models trained with datasets with $N_{ppd}=20, 10$ and 5, respectively.
    The half-filled points threaded by the black dash line correspond to the case 
    of equal relative error and input noise (the identity line, 
    which would be diagonal in a linear plot). 

    In general the error statistics appear to trace the value of the 
    pre-regularization noise.
    The RMS value remains well below the identity line, 
    consistent with the quality of the gradient data obtained 
    after Tikhonov's regularisation, except at the low noise end, $\sigma_n\leq 10^{-2}$.
    Since panels on the same row correspond to prediction errors computed using the 
    same testset, their comparison shows that models trained with larger datasets, 
    i.e. with larger $N_{ppd}$ are significantly more accurate.
    This is shown more specifically in Fig.~\ref{Fig8} where the statistics
    presented in the top three panels of Fig.~\ref{Fig7}, relative to the PR-5 domain,
    are plotted as a function of $N_{ppd}$ in three separate panels.
    Although the data points are not aligned along straight lines, 
    there appear to be a trend for the various statistics to decrease linearly
    with the parameter $N_{ppd}$.
    Related to the improvement of the model performance as we move along the 
    rows of Fig.~\ref{Fig7} from right to left, which is attributed to a higher 
    sampling density of training data, there is a 
    corresponding performance worsening as we compare panels from top to bottom 
    along vertical columns which, 
    in addition to the reverse of the above effect,
    includes the additional difficulty of modeling the
    function's gradient on a progressively larger domain.
    
    \subsection*{Convergence Tests {\bf [NEW]}}
    To test the performance of our MLP representation 
    in a numerical context, we set out to compute the temperature
    evolution of a plasma in a one-dimensional domain.
    The thermodynamic state of the plasma is constant in space, 
    except for a sinusoidal perturbation 
    with 5\% amplitude in both the temperature and a second randomly 
    chosen variable ($n$ or $T$).
    The temperature evolution is computed by time integrating the  
    equation \ref{eq:cont},
    using a second order accurate numerical scheme for hyperbolic
    equations\cite{vanLeer1977}
    that employs the heat flux gradient provided by our MLPs.
    The algorithm, further detailed in 'Numerics'
    in Methods, is a higher order extension of Godunov
    method using a predictor corrector scheme.
    The computational domain has 
    periodic boundary conditions and is discretised with $N_{\rm Mesh~ Points}$ resolution elements.
    The performance is based on the convergence rate
    of the numerical solution, i.e. the rate at which the error drops 
    as a function of $N_{\rm Mesh~ Points}$. The numerical
    error is computed using Richardson's 
    extrapolation method (also further detailed in 'Numerics' in Methods).
    We repeat the numerical integration test for a sample of 30 runs 
    with different, randomly chosen, unperturbed ($n, T, \beta$) plasma parameters.

    The results are summarised in Fig. \ref{Fig9} where, 
    for each MLP model, labeled according to the noise characterising the unregularised heat-flux data, the sample averaged $L_2$ (blue dashed line) and $L_{\infty}$
    (red dashed line) error norms
    are plotted as a function of, $N_{\rm Mesh~ Points}$.
    The corresponding results for the analytic form of the heat-flux function
    are also shown (gray and cyan dashed line for $L_2$ and $L_\infty$ respectively)
    together 
    the $L_2$ error of the initial conditions (olive dashed line),
    providing a simple sanity check.
    The $L_2$ and $L_\infty$ errors norms of 
    the various MLP based integration scheme implementations
    actually overlap and, more importantly, display a second order
    convergence rate (see black line),
    as expected for a second order accurate scheme.
    The slight flaring of $L_\infty$ at the high resolution end is 
    sensitive to the sinusoidal amplitude and 
    is perhaps indicative of nonlinear effects.
    The implementation using the analytic heat-flux function
    shows the same convergence - actually this test simply verifies the 
    correctness of our code implementation - as would 
    any other analytic expression, including the results of 
    our regression analysis.
    Notice that the numerical solutions corresponding to each 
    implementation do converge to different results. 
    This is a consequence of the consistency property of the scheme\cite{Allaire2007}.
    In fact, in each case
    we are modeling a slightly different equation 
    specified by the specific flux function model (analytic, MLP or symbolic regression)
    utilised in the numerical scheme.
    
    The second order convergence of the implemented scheme confirms the
    error analysis presented earlier in equation~(\ref{eq:erranalysis}).
    To further illustrate that it is the stochastic oscillation of the gradient
    in the neighbor of an expansion point that causes the random error cancellation
    of the spurious term 
    and the resulting partial improvement of the convergence rate with respect to 
    the expectation from the truncation error, 
    we carry out the following simple experiment.
    We implement the above integration scheme
    using the flux model in equation (\ref{eq:stochgrad}) and compare the results 
    obtained with three different models for $a$:
    in addition to the original 
    `stochastic' value sampled between [-1, 1]
    we also set $a=0$ (i.e. the `analytic' model) and $a=1$ (`fixed' value).
    We carry out the same numerical integration tests as above except that we
    now perturb only the temperature component.
    The results for the three models are shown in Supplementary Fig.~\ref{FigA3} 
    together with the usual ICs sanity check.
    The plot shows that while the `stochastic' model (gray)
    displays second order convergence rate (see bottom black line)
    as the the `analytic' model (blue), 
    the `fixed' model (red) converges only as $N^{-1.5}$.
    This would be expected for a consistent error build up 
    due to the spurious fixed term during each time-step integration.

    \subsection*{Symbolic Regression}
    In this final stage we attempt to recover a mathematical expression for
    the heat flux function through a symbolic regression analysis.
    For this purpose we use the Deep Symbolic Optimisation package~\cite{Petersen2021} 
    which builds the mathematical expressions through a 
    recurrent neural network trained with a reinforcement learning method.
    We first, however, precondition our symbolic regression 
    in two ways: first, as suggested 
    in~\cite{Udrescu2020}, we restrict the search to expressions with sensible physical dimensions.
    In addition, we exploit knowledge of the asymptotic limit of the sought mathematical expression if known.
    This is common practice when seeking to extend a law of physics to previously unexplored 
    regimes (e.g., from classic to the relativistic or quantum limits).
    Both of these measures help reduce the symbolic search space which grows exponentially with the number of components.
    As already mentioned, the heat flux is well known when the electron mean free path is small compared to the temperature gradient scale, i.e.
    \begin{equation} \label{eq:asympt}
        \lim_{\lambda_{ei}/L_T \to 0}  q(n_e,T_e,B_e) = q_* \,\frac{\lambda_{ei}}{L_T}, \quad q_*=n_eT_ev_{te}.
    \end{equation}
    Since $q_*$ has the same physical dimensions as $q$, we only need to 
    search for a 
    multiplicative dimensionless factor, our $\epsilon$ in equation~(\ref{eq:qkom}), which will be a 
    function of dimensionless variables.
    Given the dimensional physical quantities entering our plasma physics problem, 
    $(n_e,T_e,B_e,m_e,e,L_T)$, only three dimensionless combinations are possible (or combinations 
    thereof) namely, $L_T n_e e^4 T_e^{-2} \propto L_T/\lambda_{ei} \equiv x_1$, which in fact already
    appears in the asymptotic limit \ref{eq:asympt}, $n_eT_e/B^2 \propto \beta_e \equiv x_2$, and, 
    $n_e^{1/3}L_T\equiv x_3$, which actually 
    $\epsilon$ does not depend upon. Notice that from the point of view of the symbolic regression there is no advantage in 
    choosing $x_2=\beta_e$ or $x_2= n_eT_e/B^2$, or $x_1$ versus $3x_1$ for that matter, because the analysis will have to figure out the value of those coefficients by itself. Instead, we chose 
    $\beta_e$ and $L_T/\lambda_{ei}$ because they 
    have a clear physical meaning.

    The datasets for the symbolic regression consist of 2000 entries
    containing the values of the target function, $\epsilon\equiv q/q_*$ 
    and the corresponding independent variables $(x_1, x_2,  x_3)$.
    The entries are randomly sampled from the four Datasets A.1, A.5 
    A.10 and A.20 in Table~\ref{tab:data}, allowing us to compare the 
    performance of the symbolic regression under various conditions of
    data quality.
    Our function set includes a minimal choice of
    $\{+, -,\times, \div, {\rm const.} \}$, with a max of three constants,
    as well as the functions $\log$ and $\exp$, allowing 
    for generic exponential expressions, often seen in physics,
    like, '$\exp(g(\log(x))$', where $g$ is an 
    arbitrary combination of the function set.
    We also avoid 
    trigonometric functions, which are not expected in this problem.
    We set ${\tt batch\_size}=10^4$ and ${\tt n\_samples}=10^6$,
    resulting in 100 iterations and use standard settings otherwise. 
    The chosen number of iterations appears sufficient for the 
    reinforcement learning's reward function to reach a plateau, but 
    is otherwise
    arbitrary and longer runs could lead to slightly better accuracy.
    The DSO optimizes an objective function given by the
    Normalised Root Mean Squared Error (i.e. the root of the MSE
    of the relative error) of the function prediction, similar
    to the case of the previously discussed regularisation and MLP training.
    
    {\bf
    The results are summarised in Table~\ref{tab:symbolic} where for each 
    dataset we report the symbolic expression obtained through the regression,
    the RMS and Max of the relative error for both the flux and its gradient,
    computed on a random sample of 10$^6$ entries.
    The symbolic regression appears to successfully retrieve the 
    correct functional form of the factor $\epsilon$,
    clearly outperforming the MLP model
    in terms of uncertainties of the flux gradient.
    Exception is made for 
    the case of the A.5 Dataset, in which the symbolic formula is
    found to be characterised by an unusually large error.
    This result is rather peculiar and
    illustrates the importance of running the regression with
    multiple random initialisation seeds to obtain statistically
    robust results.
    The other interesting observation is that there seems to be no obvious 
    advantage from running the DSO on the regularised data.
    As in the case of the MLP models, 
    this seems related to the non Gaussian character
    of the residual error in the regularised data, which the
    NRMSE minimization carried out by the DSO is not effective at reducing.
    In this respect it is interesting to compare the statistics of residual errors after
    Tikhonov's regularization given in Supplementary Fig.~\ref{FigA2} and those of the
    symbolic regression in Table~\ref{tab:symbolic}, and notice, 
    with a grain of salt, their comparable magnitude.

    In conclusion, it is difficult to predict the performance of the DSO
    in the case of more complex functional dependencies 
    between features and labels, and when the statistics 
    of the data errors is not purely Gaussian.
    Nevertheless, the performance shown in a study of 
    realistic research interest even with high levels 
    of data noise is encouraging,
    particularly from the perspective of using experimental data.
    }

\section*{Conclusions [{\bf NEW}]} 
\label{sec:conclusions}
    In this paper we use a ML based approach to
    improve basic knowledge, mathematical description and 
    numerical modeling capability of generic transport processes. 
    Ours is part of ongoing efforts to employ modern artificial intelligence 
    techniques in science and overlaps in scope with
    developments of augmented schemes and accelerators for 
    numerical simulations, 
    as well as methods to gain insight in and possibly reach 
    discovery of new laws of physics through symbolic regression analysis.
    Transport processes may be ruled by 
    complex micro-physics which is impractical to model
    theoretically, but may exhibit emergent behavior 
    describable by a closed mathematical expression.
    We are, therefore, particularly interested in formulating transport
    terms, ${q}$, employable in a continuum mechanics macroscopic description,
    by learning from data provided either by 
    microscopic-scale numerical simulations, progressively 
    closer to first principles, or even directly from experiments,
    for those physical conditions under which even current
    {\it ab initio} codes do not provide consistent results.
    
    Ideally one would be able to learn the transport term, ${q}$, 
    as a mathematical expression obtained through a symbolic regression analysis.
    This is the preferred path because it allows for easier 
    implementation in computational modeling and 
    is very valuable to theoretical analysis. However, 
    it is also the less certain path, 
    because it is an intrinsically more
    difficult task due to the unknown complexity of the
    sought relation and, amongst others, the degrading impact
    of the data uncertainties on its performance\cite{Udrescu2020}.
    Alternatively, the transport term, ${q}$, can be expressed via an MLP representation.
    Care must be taken, however, with its implementation in numerical integration 
    codes.
    Due to its noisy character, its Taylor expansion is unreliable 
    beyond the first order term, leading to truncation errors 
    $\tau (\Delta x) \sim O(\Delta x^{1/2})$.
    We have nevertheless shown that owing, in particular,
    to the peculiar stochastic character of the
    error term spoiling the Taylor expansion,
    using an MLP representation of
    $\nabla q$ instead of ${q}$ itself,
    allows us to effectively recover a flux function sufficiently smooth for 
    implementation in a second order accurate code.
    Formulations for even higher order schemes are feasible but 
    add complexity and were not explicitly pursued here.
    In any case, in this approach
    it becomes necessary to first regularize the flux data,
    to minimize the impact of the error noise on the 
    calculation of the flux gradients, which
    provide the labels for our trainable MLP function.
    Our method of choice for this purpose is Tikhonov's.
    
\begin{figure}[ht]
        \centering
        \includegraphics[width=1.0\textwidth,]{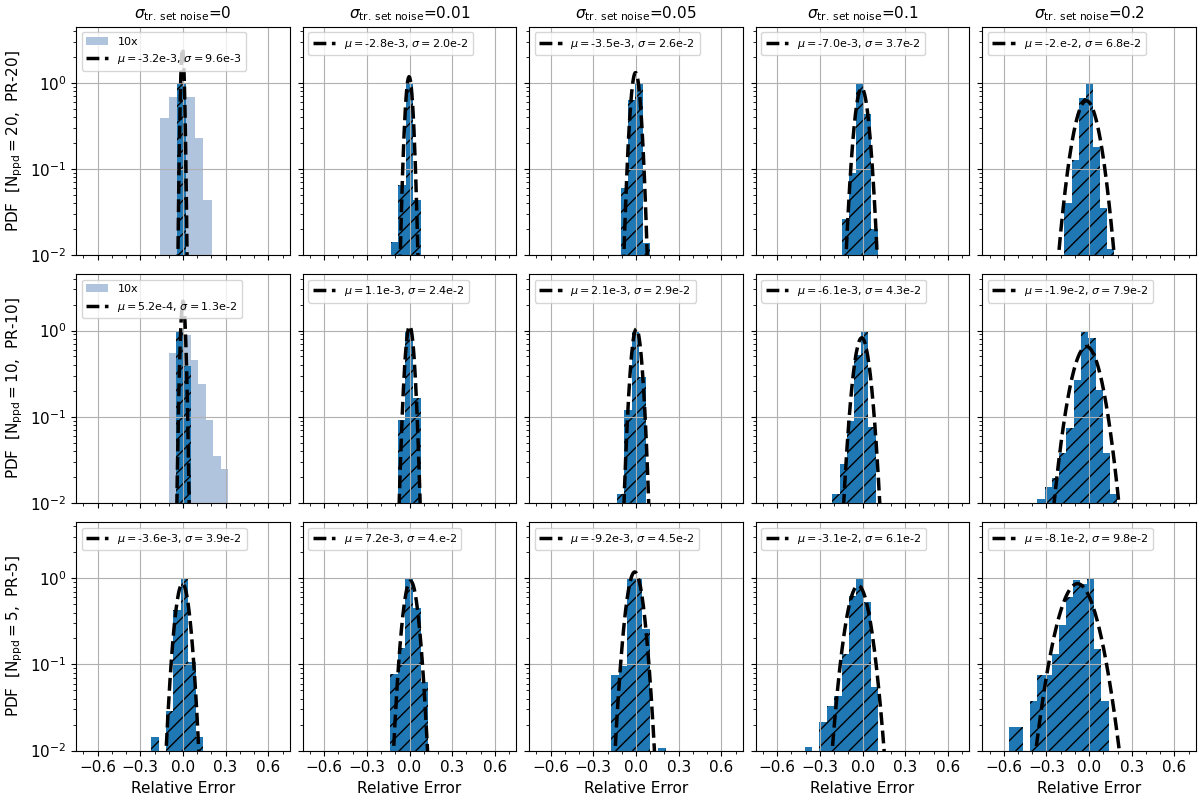}
        \caption{MLP models test errors:
        Histogram of the {\it test-errors} of the MLP models in Table~\ref{tab:models}. 
        All the histograms are rescaled so that they all peak at 1.
        Each row corresponds to a different $N_{ppd}$  
        (and parameter range domain PR-$N_{ppd}$), while each column corresponds to a 
        different value of $\sigma_n$, the noise in the flux dataset before applying
        Tikhonov's regularization.
        The errors are computed with respect to the noiseless testsets, i.e.
        the A.0, B.0 and C.0 testsets for models of the A-, B-, C- series, respectively.
        The legend shows the mean $\mu$ and standard deviation $\sigma$ of the histogram
        in each panel.
        The lightblue shapes in some panels correspond to a histogram of 10 $\times$ larger errors.
        \label{Fig6}
        }
    \end{figure}

\begin{figure}[ht]
        \centering
        \includegraphics[width=0.9\textwidth,]{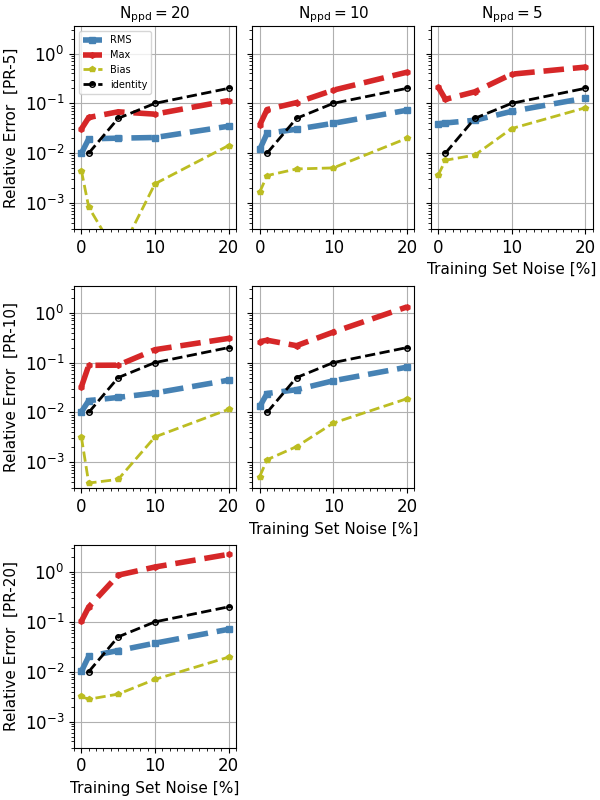}
        \caption{Test-error Statistics: 
        RMS (blue dash line), Max (red dash line) and Bias (yellow thin dash line)
        statistics of the model prediction errors
        and their trend with the pre-regularization noise
        for different sampling size cases.
        In particular, from top to bottom, the rows correspond to errors computed using 
        the noiseless testset of the C-, B- and A- Series respectively,
        while from left to right the columns
        correspond to models trained with datasets with $N_{ppd}=20, 10$ and 5, respectively.
        The half-filled points threaded by the black dash line correspond to the case 
        of equal relative error and input noise (the identity line, 
        which would be diagonal in a linear plot). 
        \label{Fig7}
        }
    \end{figure}

\begin{figure}[ht]
        \centering
        \includegraphics[width=1.0\textwidth,]{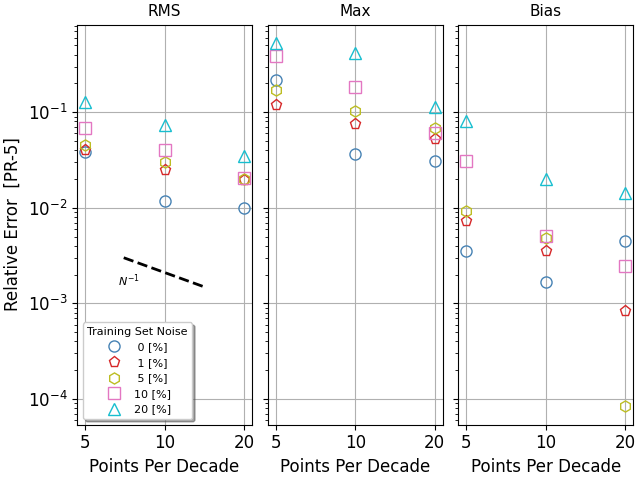}        
        \caption{Trend with data sampling density:
         Test-errors' RMS (left), Max (center) and Mean (left) presented
         in the top three panels of Fig.~\ref{Fig7}, relative to the PR-5 domain,
         plotted as a function of $N_{ppd}$.
         Different symbols correspond to models trained with data characterised 
         by different pre-regularization noise (see Figure's legend).
         The error statistics appear to roughly decrease as the inverse 
         of the parameter $N_{ppd}$ (black dashed line).
        \label{Fig8}}
    \end{figure}

\begin{figure}[ht]
        \centering
        \includegraphics[width=1.0\textwidth,]{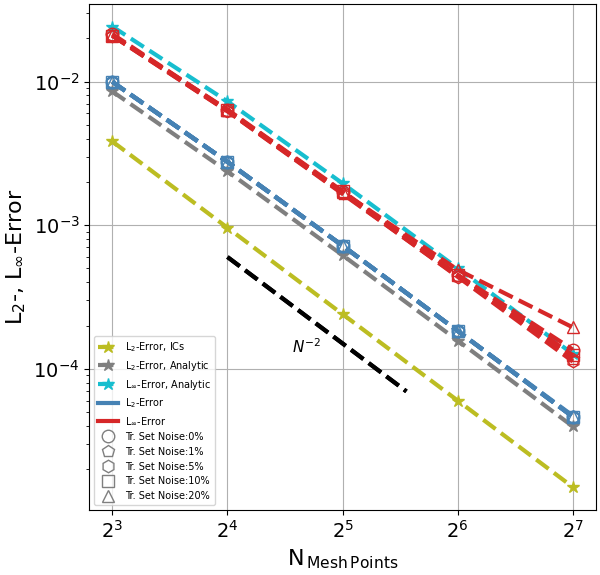}        
        \caption{Convergence test:
        $L_2$ and $L_\infty$ error norms 
        for the implementations using an MLP model of the flux gradient
        (blue and red, respectively)
        and for the implementation using instead an analytic expression
        (gray and cyan, respectively, with the cyan curve multiplied by 1.15
        to make it visible).
        The plotted errors are averages over a sample of 30 runs using 
        different, randomly chosen, unperturbed  values of the 
        thermodynamic parameters, ($n, T, \beta$).
        Models trained with data characterised by different pre-regularization 
        noise are represented by different symbol (see legend), 
        though they are difficult to distinguish as their calculated error 
        data points mostly overlap.
        The $L_2$ error norm is also shown for the initial conditions (olive).
        Expected error drop rate for a second order accurate scheme
        is indicated by the black dash line.
        \label{Fig9}}
    \end{figure}

    When applied to an idealised study of heat transport relevant to
    astrophysical and thermonuclear fusion plasmas we find that
    Tikhonov's regularization is very effective at cleaning the data from Gaussian 
    noise, allowing accurate estimates of the flux gradient.
    The MLP's trained on such labels deliver in general flux gradient representations
    of relatively high quality, with their overall performance that, while reflecting
    the pre-regularization noise level, appears to improve roughly linearly
    with the density of parameter sampling (Fig. \ref{Fig8}).
    For example, for a relative error noise of 10\% and $N_{ppd}=20$ our MLP 
    representation computes the flux gradient components with an error RMS
    of 3.7\% (Fig. \ref{Fig6}).
    Interestingly, the MLP training based on regularised data does not lead to model
    prediction errors that are further reduced with respect to the error 
    characterising the training data.
    This is clear when comparing the error RMS of the
    regularised gradient data 
    in Supplementary Fig. \ref{FigA2} and of the MLP gradient predictions in Fig. \ref{Fig6}.
    In particular, for each MLP representation, at the low-end of the $\sigma_n$ values
    we have a prediction error RMS $\simeq\sigma_n$ while at the high-end,
    i.e. $\sigma_n =10, 20$, we have a prediction error RMS $\simeq \sigma_n /2
    \simeq 2 \sigma_{reg}$, i.e. half the pre-regularizaton noise
    but twice as large the error
    of the regularised data (Supplementary Fig. \ref{FigA2}).
    We believe this is caused by the non Gaussian character of the residual 
    error of the regularised data which makes the MSE a less effective objective
    for closing in on the ground truth.
    Finally, the latent representation defined by our MLPs is shown to be suitable for 
    implementation in second order accurate schemes and 
    can be used in computational models of the continuity equation
    as a plug-in to augment the
    numerical integration algorithm and deploy and accurate
    description of the transport process of interest.

    The application of the DSO symbolic regression package to our 
    idealised study of heat transport, complemented  
    by some preconditioning operations of the problem,
    leads to a successful results
    mostly outperforming the precision of the MLP model,
    though it may be useful to run the regression with
    multiple random initializations to help avoiding occasional 
    lack of performance.
    In line with the findings related to the MLP results,
    we observe that running the DSO on the regularised data
    appears to produce little or no benefit.
    This seems again related to the non Gaussian character
    of the regularised data's error which the
    DSO optimization based on the NRMSE 
    is not so effective at reducing.
    In this respect, we note the comparable 
    performance of Tikhonov's regularization in Supplementary Fig.~\ref{FigA2} and
    the DSO prediction in Table~\ref{tab:symbolic} in terms of error statistic.
    
    The foregoing discussion suggests that while it is preferable to have accurate data
    it is also important to have control over the error statistics
    in order to employ the most appropriate objective function.
    Along the same line, we expect our results to remain valid 
    beyond the specific case of random Gaussian relative error assumed in
    our idealised study, provided the objective functions are modified accordingly.

    In conclusion, the successful retrieval of accurate MLP 
    and symbolic representations of the heat flux function 
    in the context of a study that is somewhat idealised, 
    yet representative of the field\cite{Komarov2018},
    appears a promising first step to be able to employ 
    in macroscopic models the necessary 
    sophisticated information on transport processes implicitly
    available from microscopic descriptions.
    The robustness of the results even in the case of 
    significant noise in the data is also very attractive,
    particularly in view of applications using
    experimental data.

\section*{Methods}
    \subsubsection*{MLP Architecture}
     The architecture of our MLP is summarised in Fig.~\ref{Fig3}.
     At its core is a number, $N_{Layers}$, 
     of hidden layers each with 
     the same number, $N_{Units}$, of hidden units, both tunable parameters.
     We embed the input features into a set of
     Random Fourier Features~\cite{RFF,Tancik2020} (RFF), i.e. given the input 
     vector $\mathbf{x}$, we define the components
     \begin{equation}
         \mathbf{x}_i \leftarrow \cos(\mathbf{k}_i\mathbf{\cdot x +  \phi}_i), \hspace{0.5cm}
         i\in \{i: 0\leq N_{\rm RFF} \}.
     \end{equation}
     with $N_{\rm RFF}=N_{Units}$.
     As in \cite{Tancik2020} we observe no benefit when training the parameters $\mathbf{k}_i$ 
     and $\phi_i$, so following \cite{RFF} we randomly sample the $\mathbf{k}_i$'s from the 
     distribution,  $\mathcal{N}(0,\sigma_{\rm RFF})$, with $\sigma_{\rm RFF}$ a tunable 
     parameter, and the $\phi_i$'s uniformly in the interval $[0, 2\pi)$.
     A ReLU activation function is applied to the affine mapping returned by the hidden units. We also employ skip connections to feed the RFF embeddings to every other hidden layer except the last. The output layer consists of as many regression units as the gradient component
     without activation function.
     To train the MLP we define a loss function given by the Mean Squared Error of the predicted value with respect to the label.
     To prevent overfitting we early stop the training if the accuracy does not improve during 
     a number of consecutive iterations given by a patience parameter set to 100.

    \subsubsection*{MLP Representation of the Flux Function {\bf [NEW]}}
    The MLP representation of the flux function uses
    the architecture described in the previous section with the following custom
    choice of the otherwise tunable parameters: 5 hidden layers each consisting of 512 units,
    a $\sigma_{\rm RFF}=0.95$ for RFF embedding and a $L_2$ 
    regularization with parameter $\lambda=10^{-5}$.
    The MLP was trained with the noiseless data of the A.0 set using a learning 
    rate of $10^{-3}$. It reached convergence after 410 integration
    steps resulting in an RMS error of $9.6\times10^{-3}$ and a
    MAX error of $3.8\times10^{-2}$.
    
    \subsubsection*{Regularization {\bf [NEW]}}
    In this section we develop the tools to compute $f$'s derivative
    of order $p$, $f^{(p)}$, based on finite difference schemes. This are the tools 
    employed for the calculation of the regularization term in Tikhonov's method. 

    We consider a $D-$dimensional parameter space discretized by a grid 
    $\Xi\in\mathbb{R}^D$ of dimensions $n_0,\dots n_{D-1}$
    and a scalar function $f\colon {\bf \xi} \in \Xi\longrightarrow\mathbb{R}$.
    The grid elements, ${\bf \xi}$, identified by the set of indexes $i_0, i_1, \dots i_{D-1}$,
    are not necessarily uniformly spaced but their parameter values
    grow monotonically with the respective index.
    In view of what will follow we define the utility function
    \begin{align*}
    \mathcal{V}_{{\tt cmp}, \,d} \equiv \begin{cases}
    \Pi_{i \in A \coloneqq \{i \;\mid\; 0\leq i < D \,\land \, i \,{\tt cmp} \,d\}} \; n_i, \quad &{\rm if \quad A \neq \varnothing} \\
    1 \quad &{\rm otherwise}
    \end{cases}
    \end{align*} 
    which takes as input a comparison operator, 
    ${\tt cmp}$, and a dimension, $d$, and returns 
    the volume of the subspace consisting of the dimensions fulfilling 
    the comparison. For example, $\mathcal{V}_{>,\, 1}$ returns $n_2\times n_3\times\dots n_{D-1}$. 
    We also define a stacking function 
    \begin{equation} \nonumber
    {\mathcal S}
    (\mathcal{A}^{q}, i, n)\equiv
    \begin{pmatrix}
    \mathcal{A}^{i}, \mathcal{A}^{i+1}, \cdots , \mathcal{A}^{i+n-1}
    \end{pmatrix} ^T
    \end{equation}
    that takes an indexed operator, $\mathcal{A}^q$,
    an initial index value, $i$, and a count, $n$, and returns 
    a stack of $n$ contiguously indexed operators starting from index $i$.
    In order to proceed we now linearise the grid of parameter values in $\Xi$
    by arranging them into a 1-dimensional array, $\mathcal{X}$, according 
    to an order in which
    the highest index runs fastest and the lower indexes progressively slower.
    Likewise we define a 1-dimensional array, ${\bf y}$, whose $i$-th element 
    is $y_i=f({\bf x}_i)$, with ${\bf x}_i$ the $i$-th element of $\mathcal{X}$.

    We can now define the matrix operator, 
    $\Delta^{1,\,m,\,d}_{n_0,\dots n_{D-1}} 
    \colon \mathbb{R}^{n_0\cdot n_1\dots n_{D-1}} \longrightarrow
    \mathbb{R}^{n_0\cdot n_1\dots (n_d-m)\dots n_{D-1}} $, 
    computing the difference between values corresponding to grid
    points separated by $m$ points along the $d$ axis,
    \newpage
    \begin{strip}
    \begin{eqnarray*} \nonumber
    {\Delta}^{1,\,m,\,d}_{n_0,\dots n_{D-1}}= 
    \begin{matrix}
    \begin{pmatrix}
    {\mathcal S}(\hat{\Delta}^{m,\,d}_{n_0,\dots n_{D-1}}, 0, \mathcal{V}_{>d}) &
    \makebox[\dimen2]{\Large$0$} & \cdots & \makebox[\dimen2]{\Large$0$} \\[0.75em]

    \makebox[\dimen2]{\Large$0$} & 
    {\mathcal S}(\hat{\Delta}^{m,\,d}_{n_0,\dots n_{D-1}}, \mathcal{V}_{>d}, \mathcal{V}_{>d}) &
    \cdots & \makebox[\dimen2]{\Large$0$}  \\[0.75em]

    \vdots & \vdots & \ddots & \vdots \\[0.75em]
    \makebox[\dimen2]{\Large$0$} & \makebox[\dimen2]{\Large$0$} & \cdots &
    {\mathcal S}({\hat{\Delta}}^{m,\,d}_{n_0,\dots n_{D-1}},\mathcal{V}_{\ne d}-\mathcal{V}_{>d}, \mathcal{V}_{>d})
    \end{pmatrix}
    \end{matrix}
    \end{eqnarray*}
%
    where
%
    \begin{eqnarray*} \nonumber
    \hat{\Delta}^{m,\,d,\, k}_{n_0,\dots n_{D-1}} =
    \begin{matrix}
    
     \begin{matrix}
     \hspace*{-0.5\dimen0}  
     \topA 
     \hspace*{0.5\dimen0}
     \end{matrix} \\[0.75em]

    \begin{pmatrix}
    \overbrace{0\cdots 0}^{k} & -1 & \overbrace{0\cdots 0}^{m\cdot \mathcal{V}_{>d}-1} & 1  & 0 & 0\cdots 0  &  \overbrace{0\cdots 0}^{\mathcal{V}_{>=d} -k-1}  \\
    0\cdots 0 & \overbrace{0\cdots 0}^{\mathcal{V}_{>d}} & -1& 0\cdots 0 & 1 & 0\cdots 0 &  0\cdots 0 \\
    \vdots & \vdots & & \ddots &  &\vdots &\vdots  \\
    0\cdots 0 & 0\cdots  & \cdots 0 
    & -1 & \overbrace{0\cdots 0}^{m\cdot \mathcal{V}_{>d}-1} & 1 & 0\cdots 0 \\
    \end{pmatrix}
    \hspace*{.15\dimen0} \begin{matrix} \rightbrace \end{matrix}
    \\
    \begin{matrix} \hspace*{-3.9\dimen0}   \bottomC \end{matrix}
    \end{matrix} .
    \end{eqnarray*}
    \end{strip}

    The partial differential operator with respect 
    to the $d$-th component of ${\bf x}$ is then
    \begin{align*}
    \mathcal{D}^{1,\,m,\,d}_{n_0,\dots n_{D-1}} = 
    {\tt Diag}^{-1}(\Delta^{1,\,m,\,d}_{n_0,\dots n_{D-1}} \cdot {\bf x}) \cdot
    \Delta^{1,\,m,\,d}_{n_0,\dots n_{D-1}}
    \end{align*}
    and
    \begin{align*}
    {\bf y}^{(1)} =
    \mathcal{D}^{1,\,m,\,d}_{n_0,\dots n_{D-1}} \cdot {\bf y}
    = \frac{\Delta^{1,\,m,\,d}_{n_0,\dots n_{D-1}}\cdot {\bf y}}{\Delta^{1,\,m,\,d}_{n_0,\dots n_{D-1}} \cdot {\bf x}},
    \end{align*}
    with the fraction in the last term meant component-wise.
    Likewise we can define the $p$-th order finite difference operator
    with respect to the $d$-th component of ${\bf x}$, 
    \begin{align*}
    {\Delta}^{p,\, m, \, d}_{n_0,\dots n_{D-1}}= \\ 
    \overbrace{
    {\Delta}^{1,\, m, \, d}_{n_0,\dots n_d-m*(p-1), \dots n_{D-1}} \cdot
    {\Delta}^{1,\, m, \, d}_{n_0,\dots n_d-m*(p-2), \dots n_{D-1}} \cdot 
    \dots {\Delta}^{1,\, m, \, d}_{n_0,\dots n_{D-1}}}^{p \;{\rm times}}
    \end{align*}
    and the corresponding $p$-th partial differential operator
    \begin{align*}
    \mathcal{D}^{p,\,m,\,d}_{n_0,\dots n_{D-1}} = 
    {\tt Diag}^{-1}(\Delta^{p,\,m,\,d}_{n_0,\dots n_{D-1}} \cdot {\bf x}) \cdot
    \Delta^{p,\,m,\,d}_{n_0,\dots n_{D-1}}
    \end{align*}
    so that 
    \begin{align*}
    {\bf y}^{(p)} = \mathcal{D}^{p,\,m,\,d}_{n_0,\dots n_{D-1}} {\bf y} = 
    \frac{\Delta^{p,\,m,\,d}_{n_0,\dots n_{D-1}} \cdot {\bf y}}{\Delta^{p,\,m,\,d}_{n_0,\dots n_{D-1}} \cdot {\bf x}}.
    \end{align*}

    For $m=1$, the above operators map grid point values to midpoint interface values 
    and for $m=2$ it maps grid point values to interior midpoint grid values.
    If $m$ is even and the grid spacing is uniform, ${\bf y}^{(p)}$ and ${\bf x, ~y}$ 
    share the same $\Xi$ grid except for an outer layer of thickness $m$,
    as finite differences cannot be computed normal to the boundary
    unless proper boundary conditions are provided.
    The above differentials can also be composed to build mixed differentiation.
    Finally, differential operators of various order and compositions 
    can be stacked to define an overall matrix operators 
    providing the regularization term in Tikhonov's method.
    For example, a list of all partial derivatives of order $p$ (excluding the mixed terms)
    is obtained by applying to the input vector
    ${\bf y}$ the operator obtained after stacking the individual
    $\mathcal{D}^{p,\,m,\,d}_{n_0,\dots n_{D-1}}$
    for $d=0,1,\dots D-1$, as follows
    \begin{equation} \nonumber
    \mathcal{D}^{p,\,m,\,*}_{n_0,\dots n_{D-1}} = 
    {\mathcal S}(\mathcal{D}^{p,\,m}_{n_0,\dots n_{D-1}}, 0, \,n_D).
    \end{equation}

    \subsubsection*{Numerics {\bf [NEW]}}
    Our numerical integration scheme 
    is a simplified version of the
    predictor corrector scheme proposed by van Leer's\cite{vanLeer1977}.
    It is described in the pseudocode Algorithm~\ref{alg:vanleer},
    where $u=(n, T, \beta)^{\rm T}$ denotes the set of thermodynamic variables,
    $\Delta x$ and $\Delta t$ are the mesh and time-step size, 
    $N_x$ and $N_t$ the numbers of meshes and integrations time-steps, respectively,
    P.B.C. stands for application of periodic boundary conditions necessary
    to complete the operations in the next code block.
    In addition JF is formally the Jacobian of the 
    vector function whose components
    describe the flux of each thermodynamic variable. 
    Since in this case only
    the heat-flux $q$ is non-zero we have
    \begin{align*}
    {\rm JF}\colon u & \longrightarrow 
    \begin{pmatrix}
    0 & 0 & 0 \\
    \partial_n q(u) & \partial_T q(u) & \partial_\beta q(u) \\
    0 & 0 & 0
    \end{pmatrix}.
    \end{align*}
    where $\partial_z$ denotes the partial derivative with respect to $z$.
    Note that we do not use slope limiters.

    The experiments reported in the `Convergence Tests' are 
    characterised by the following setup:
    \begin{align*}
    N_x&=\left\{2^3, 2^4, 2^5, 2^6, 2^7, 2^8 \right\} \\
    \Delta x&=\frac{1}{N_x}, \\
    \Delta t&={\rm CFL}\,\frac{\Delta x}{\lambda_{max}}, \\
    N_t &= 2 \,\frac{N_x}{2^3}.
    \end{align*}
    with CFL=0.5 and $\lambda_{max}$ the max value over the domain
    of the (only) eigenvalue of the problem, namely $\lambda=\partial_T q$,
    thus enforcing the CFL condition on the timestep.
    
    The errors are measured using Richardson's extrapolation.
    So, given the numerical result $T_{r}$ at a given resolution $r$ 
    we first estimate the error at a given grid point $i$, as
    \begin{align*} 
    \varepsilon_{r, i} = T_{r,i} - \bar T_{r+1,i},
    \end{align*}
    where $\bar T_{r+1}$ is the solution at the next finer resolution,
    spatially averaged onto the coarser grid (which is second order 
    accurate). We then take the 2-norm and max-norm of the error,
    \begin{align*} 
    L_2 & = \| \varepsilon_{r} \|_2 =  \left( \sum |\varepsilon_{r,i}|^2  v_{i}\right)^{1/2},\\
    L_{\infty} & = \| \varepsilon_{r} \|_\infty = \max(|\varepsilon_{r,i}|) \label{lnorm_n:eq}
    \end{align*}
    where, $v_{i}=\Delta x$ is the cell volume.

    \begin{algorithm}
    \renewcommand{\algorithmicrequire}{\textbf{Input:}}
    \renewcommand{\algorithmicensure}{\textbf{Output:}}
    \caption{Simplified van Leer's predictor-corrector integration scheme.}\label{alg:vanleer}
    \begin{algorithmic}
    \Require{$u^0_{0} \dots u^0_{N_x-1}$, $\Delta x$, $\Delta t$, $N_x$, $N_t$, JF}
    \Ensure{$u^{N_t}_{0} \dots u^{N_t}_{N_x-1}$}
    \State $n \gets 0$
    \While{$n < N_t$}

    \State $u^{n}_{i}$ $\gets$ $u^n_i$ + P.B.C

    \For{$i \gets 0$ to $N_x-1$}
        \State $\delta u^n_{i+\frac{1}{2}}$ $\gets$ $u^n_{i+1}-u^n_{i}$
        
        \State $\left(\frac{\Delta F}{\Delta u}\right)^n_{i+\frac{1}{2}}$ $\gets$ 
        JF$\left( u^n_i + \frac{1}{2} \delta u^{n}_{i+\frac{1}{2}} \right)$

        \State $\tilde u^{n+\frac{1}{2}}_{i+\frac{1}{2}}$ $\gets$ $u^n_i+ \frac{1}{2}\,
        \left(1-\frac{\Delta t}{\Delta x} \left(\frac{\Delta F}{\Delta u}\right)^n_{i+\frac{1}{2}} \right) \,\delta u^n_{i+\frac{1}{2}}$
    \EndFor

    \State $\tilde u^{n+\frac{1}{2}}_{i+\frac{1}{2}}$ $\gets$ $\tilde u^{n+\frac{1}{2}}_{i+\frac{1}{2}}$ + P.B.C

    \For{$i \gets 0$ to $N-1$}
        \State $\left(\frac{\Delta F}{\Delta u}\right)^{n+\frac{1}{2}}_{i}$ $\gets$ 
        JF$\left( \frac{1}{2}\left[\tilde u^{n+\frac{1}{2}}_{i-\frac{1}{2}} + \tilde u^{n+\frac{1}{2}}_{i+\frac{1}{2}} \right] \right)$

        \State $\delta u^{n+\frac{1}{2}}_i$  $\gets$ $\tilde u^{n+\frac{1}{2}}_{i+1}-\tilde u^{n+\frac{1}{2}}_{i-\frac{1}{2}}$

        \State $u^{n+1}_i$  $\gets$ $u^n_i -\frac{\Delta t}{\Delta x} \left(\frac{\Delta F}{\Delta u}\right)^{n+\frac{1}{2}}_{i} \cdot \delta u^{n+\frac{1}{2}}_i$
    
    \EndFor
    \State $n \gets n + 1$
    \EndWhile
    \end{algorithmic}
    \end{algorithm}
    
    \subsubsection*{Hyperparameter Optimization}
    Our model is characterised by a number of hyperparameters, particularly
    the number of hidden layers and hidden units, the value of $\sigma_{\rm RFF}$, the 
    initial value of the learning rate. Appropriate range of values for these parameters
    have become clear during the development and testing stages.
    In the final stage we perform additional optimal tuning
    by comparing for each dataset listed in Table \ref{tab:data} 
    a total of 100 models with hyperparameters selected from 
    the reduced search space given in Table \ref{tab:hpsearch}.
    Other parameters not listed there include the batch size, typically set to 700, 
    and the number of steps before the decay rate of the learning rate enters into effect, 
    ranging between 800 and 1600. The hyperparameter optimisation is efficiently carried out with
    the orchestrator \code{Ray Tune}~\cite{liaw2018tune}.

    \subsubsection*{Implementation Details}
    Our code is implemented in JAX~\cite{jax2018github} and uses public
    libraries for both data-structures and algorithms.
    In particular, Tikhonov's regularisation code makes extensive use of
    SciPy libraries for sparse matrix operations, while our Deep Learning
    code is based on libraries from Deepmind including \code{Haiku}~\cite{haiku2020github} for the MLP and 
    \code{Optax}~\cite{optax2021github} 
    for the optimiser. The latter consists of a \code{chain} object 
    combining an Adam algorithm~\cite{Kingma2015} with standard settings
    and a custom exponential-decay scheduler characterised by a drop rate of
    0.9997 and a floor value of $10^{-5}$. The scheduler kicks in after an 
    input number of steps varying between 800 and 1600.

\bibliography{packgs,phys,papers}

\begin{thebibliography}{10}
\urlstyle{rm}
\expandafter\ifx\csname url\endcsname\relax
  \def\url#1{\texttt{#1}}\fi
\expandafter\ifx\csname urlprefix\endcsname\relax\def\urlprefix{URL }\fi
\expandafter\ifx\csname doiprefix\endcsname\relax\def\doiprefix{DOI: }\fi
\providecommand{\bibinfo}[2]{#2}
\providecommand{\eprint}[2][]{\url{#2}}

\bibitem{Noether1918}
\bibinfo{author}{Noether, E.}
\newblock \bibinfo{journal}{\bibinfo{title}{Invariante variationsprobleme}}.
\newblock {\emph{\JournalTitle{Nachrichten von der Gesellschaft der
  Wissenschaften zu G\"ottingen, Mathematisch-Physikalische Klasse}}}
  \textbf{\bibinfo{volume}{1918}}, \bibinfo{pages}{235--257}
  (\bibinfo{year}{1918}).

\bibitem{Leveque2002}
\bibinfo{author}{Leveque, R.~J.}
\newblock \emph{\bibinfo{title}{Finite Volume Methods for Hyperbolic Problems}}
  (\bibinfo{publisher}{Cambridge University Press},
  \bibinfo{address}{Cambridge}, \bibinfo{year}{2002}).

\bibitem{Allaire2007}
\bibinfo{author}{Allaire, G.} \& \bibinfo{author}{Craig, A.}
\newblock \emph{\bibinfo{title}{Numerical Analysis and Optimization: An
  Introduction to Mathematical Modelling and Numerical Simulation}}.
\newblock Numerical Mathematics and Scientific Computation
  (\bibinfo{publisher}{OUP Oxford}, \bibinfo{year}{2007}).

\bibitem{cottet_koumoutsakos_2000}
\bibinfo{author}{Cottet, G.-H.} \& \bibinfo{author}{Koumoutsakos, P.~D.}
\newblock \emph{\bibinfo{title}{Vortex Methods: Theory and Practice}}
  (\bibinfo{publisher}{Cambridge University Press}, \bibinfo{year}{2000}).

\bibitem{bell1981}
\bibinfo{author}{Bell, A.~R.}, \bibinfo{author}{Evans, R.~G.} \&
  \bibinfo{author}{Nicholas, D.~J.}
\newblock \bibinfo{journal}{\bibinfo{title}{{Elecron Energy Transport in Steep
  Temperature Gradients in Laser-Produced Plasmas}}}.
\newblock {\emph{\JournalTitle{Physical Review Letters}}}
  \textbf{\bibinfo{volume}{46}}, \bibinfo{pages}{243--246},
  \doiprefix\url{10.1103/physrevlett.46.243} (\bibinfo{year}{1981}).

\bibitem{Gregori2004}
\bibinfo{author}{Gregori, G.} \emph{et~al.}
\newblock \bibinfo{journal}{\bibinfo{title}{{Effect of Nonlocal Transport on
  Heat-Wave Propagation}}}.
\newblock {\emph{\JournalTitle{Physical Review Letters}}}
  \textbf{\bibinfo{volume}{92}}, \bibinfo{pages}{205006},
  \doiprefix\url{10.1103/physrevlett.92.205006} (\bibinfo{year}{2004}).

\bibitem{Brantov2013}
\bibinfo{author}{Brantov, A.~V.} \& \bibinfo{author}{Bychenkov, V.~Y.}
\newblock \bibinfo{journal}{\bibinfo{title}{{Nonlocal transport in hot plasma.
  Part I}}}.
\newblock {\emph{\JournalTitle{Plasma Physics Reports}}}
  \textbf{\bibinfo{volume}{39}}, \bibinfo{pages}{698--744},
  \doiprefix\url{10.1134/s1063780x13090018} (\bibinfo{year}{2013}).

\bibitem{Hu2016}
\bibinfo{author}{Hu, S.~X.} \emph{et~al.}
\newblock \bibinfo{journal}{\bibinfo{title}{{First-principles investigations on
  ionization and thermal conductivity of polystyrene for inertial confinement
  fusion applications}}}.
\newblock {\emph{\JournalTitle{Physics of Plasmas}}}
  \textbf{\bibinfo{volume}{23}}, \bibinfo{pages}{042704},
  \doiprefix\url{10.1063/1.4945753} (\bibinfo{year}{2016}).

\bibitem{McKelvey2017}
\bibinfo{author}{McKelvey, A.} \emph{et~al.}
\newblock \bibinfo{journal}{\bibinfo{title}{{Thermal conductivity measurements
  of proton-heated warm dense aluminum}}}.
\newblock {\emph{\JournalTitle{Scientific Reports}}}
  \textbf{\bibinfo{volume}{7}}, \bibinfo{pages}{7015},
  \doiprefix\url{10.1038/s41598-017-07173-0} (\bibinfo{year}{2017}).

\bibitem{Scudder2019}
\bibinfo{author}{Scudder, J.~D.}
\newblock \bibinfo{journal}{\bibinfo{title}{{The Long-standing Closure Crisis
  in Coronal Plasmas}}}.
\newblock {\emph{\JournalTitle{The Astrophysical Journal}}}
  \textbf{\bibinfo{volume}{885}}, \bibinfo{pages}{148},
  \doiprefix\url{10.3847/1538-4357/ab48e0} (\bibinfo{year}{2019}).

\bibitem{Komarov2018}
\bibinfo{author}{Komarov, S.}, \bibinfo{author}{Schekochihin, A.~A.},
  \bibinfo{author}{Churazov, E.} \& \bibinfo{author}{Spitkovsky, A.}
\newblock \bibinfo{journal}{\bibinfo{title}{{Self-inhibiting thermal conduction
  in a high- , whistler-unstable plasma}}}.
\newblock {\emph{\JournalTitle{Journal of Plasma Physics}}}
  \textbf{\bibinfo{volume}{84}}, \bibinfo{pages}{905840305},
  \doiprefix\url{10.1017/s0022377818000399} (\bibinfo{year}{2018}).
\newblock \eprint{1711.11462}.

\bibitem{Meinecke2021}
\bibinfo{author}{Meinecke, J.} \emph{et~al.}
\newblock \bibinfo{journal}{\bibinfo{title}{Strong suppression of heat
  conduction in a laboratory replica of galaxy-cluster turbulent plasmas}}.
\newblock {\emph{\JournalTitle{arXiv:2105.08461}}}  (\bibinfo{year}{2021}).
\newblock \eprint{2105.08461}.

\bibitem{Viellefosse1975}
\bibinfo{author}{Vieillefosse, P.} \& \bibinfo{author}{Hansen, J.~P.}
\newblock \bibinfo{journal}{\bibinfo{title}{{Statistical mechanics of dense
  ionized matter. V. Hydrodynamic limit and transport coefficients of the
  classical one-component plasma}}}.
\newblock {\emph{\JournalTitle{Physical Review A}}}
  \textbf{\bibinfo{volume}{12}}, \bibinfo{pages}{1106--1116},
  \doiprefix\url{10.1103/physreva.12.1106} (\bibinfo{year}{1975}).

\bibitem{Bernu1977}
\bibinfo{author}{Bernu, B.}, \bibinfo{author}{Vieillefosse, P.} \&
  \bibinfo{author}{Hansen, J.}
\newblock \bibinfo{journal}{\bibinfo{title}{{Transport coefficients of the
  classical one-component plasma}}}.
\newblock {\emph{\JournalTitle{Physics Letters A}}}
  \textbf{\bibinfo{volume}{63}}, \bibinfo{pages}{301--303},
  \doiprefix\url{10.1016/0375-9601(77)90910-0} (\bibinfo{year}{1977}).

\bibitem{Park2021}
\bibinfo{author}{Park, S.-H.}, \bibinfo{author}{Neeb, D.},
  \bibinfo{author}{Plyushchev, G.}, \bibinfo{author}{Leyland, P.} \&
  \bibinfo{author}{Gülhan, A.}
\newblock \bibinfo{journal}{\bibinfo{title}{{A study on heat flux predictions
  for re-entry flight analysis}}}.
\newblock {\emph{\JournalTitle{Acta Astronautica}}}
  \textbf{\bibinfo{volume}{187}}, \bibinfo{pages}{271--280},
  \doiprefix\url{10.1016/j.actaastro.2021.06.025} (\bibinfo{year}{2021}).

\bibitem{Ichimaru1962}
\bibinfo{author}{Ichimaru, S.}
\newblock \bibinfo{journal}{\bibinfo{title}{{Theory of fluctuations in a
  plasma}}}.
\newblock {\emph{\JournalTitle{Annals of Physics}}}
  \textbf{\bibinfo{volume}{20}}, \bibinfo{pages}{78--118},
  \doiprefix\url{10.1016/0003-4916(62)90117-3} (\bibinfo{year}{1962}).

\bibitem{Jezouin2017}
\bibinfo{author}{Jezouin, S.} \emph{et~al.}
\newblock \bibinfo{journal}{\bibinfo{title}{{Quantum Limit of Heat Flow Across
  a Single Electronic Channel}}}.
\newblock {\emph{\JournalTitle{Science}}} \textbf{\bibinfo{volume}{342}},
  \bibinfo{pages}{601--604}, \doiprefix\url{10.1126/science.1241912}
  (\bibinfo{year}{2013}).

\bibitem{Phillpot2005}
\bibinfo{author}{Phillpot, S.~R.} \& \bibinfo{author}{McGaughey, A.~J.}
\newblock \bibinfo{journal}{\bibinfo{title}{{Introduction to thermal
  transport}}}.
\newblock {\emph{\JournalTitle{Materials Today}}} \textbf{\bibinfo{volume}{8}},
  \bibinfo{pages}{18--20}, \doiprefix\url{10.1016/s1369-7021(05)70933-0}
  (\bibinfo{year}{2005}).

\bibitem{Qian2021}
\bibinfo{author}{Qian, X.} \& \bibinfo{author}{Yang, R.}
\newblock \bibinfo{journal}{\bibinfo{title}{{Machine learning for predicting
  thermal transport properties of solids}}}.
\newblock {\emph{\JournalTitle{Materials Science and Engineering: R: Reports}}}
  \textbf{\bibinfo{volume}{146}}, \bibinfo{pages}{100642},
  \doiprefix\url{10.1016/j.mser.2021.100642} (\bibinfo{year}{2021}).
\newblock \eprint{2108.12945}.

\bibitem{Spitzer1953}
\bibinfo{author}{Spitzer, L.} \& \bibinfo{author}{Härm, R.}
\newblock \bibinfo{journal}{\bibinfo{title}{{Transport Phenomena in a
  Completely Ionized Gas}}}.
\newblock {\emph{\JournalTitle{Physical Review}}}
  \textbf{\bibinfo{volume}{89}}, \bibinfo{pages}{977--981},
  \doiprefix\url{10.1103/physrev.89.977} (\bibinfo{year}{1953}).

\bibitem{Grabowski2020}
\bibinfo{author}{Grabowski, P.} \emph{et~al.}
\newblock \bibinfo{journal}{\bibinfo{title}{{Review of the first
  charged-particle transport coefficient comparison workshop}}}.
\newblock {\emph{\JournalTitle{High Energy Density Physics}}}
  \textbf{\bibinfo{volume}{37}}, \bibinfo{pages}{100905},
  \doiprefix\url{10.1016/j.hedp.2020.100905} (\bibinfo{year}{2020}).
\newblock \eprint{2007.00744}.

\bibitem{LeCunn2015}
\bibinfo{author}{LeCun, Y.}, \bibinfo{author}{Bengio, Y.} \&
  \bibinfo{author}{Hinton, G.}
\newblock \bibinfo{journal}{\bibinfo{title}{{Deep learning}}}.
\newblock {\emph{\JournalTitle{Nature}}} \textbf{\bibinfo{volume}{521}},
  \bibinfo{pages}{436--444}, \doiprefix\url{10.1038/nature14539}
  (\bibinfo{year}{2015}).

\bibitem{Gamahara_2017}
\bibinfo{author}{Gamahara, M.} \& \bibinfo{author}{Hattori, Y.}
\newblock \bibinfo{journal}{\bibinfo{title}{Searching for turbulence models by
  artificial neural network}}.
\newblock {\emph{\JournalTitle{Physical Review Fluids}}}
  \textbf{\bibinfo{volume}{2}}, \doiprefix\url{10.1103/physrevfluids.2.054604}
  (\bibinfo{year}{2017}).

\bibitem{Dornheim2019}
\bibinfo{author}{Dornheim, T.} \emph{et~al.}
\newblock \bibinfo{journal}{\bibinfo{title}{{The static local field correction
  of the warm dense electron gas: An ab initio path integral Monte Carlo study
  and machine learning representation}}}.
\newblock {\emph{\JournalTitle{The Journal of Chemical Physics}}}
  \textbf{\bibinfo{volume}{151}}, \bibinfo{pages}{194104},
  \doiprefix\url{10.1063/1.5123013} (\bibinfo{year}{2019}).

\bibitem{Erichson2019}
\bibinfo{author}{Erichson, N.~B.}, \bibinfo{author}{Muehlebach, M.} \&
  \bibinfo{author}{Mahoney, M.~W.}
\newblock \bibinfo{journal}{\bibinfo{title}{{Physics-informed Autoencoders for
  Lyapunov-stable Fluid Flow Prediction}}}.
\newblock {\emph{\JournalTitle{arXiv:1905.10866}}}  (\bibinfo{year}{2019}).
\newblock \eprint{1905.10866}.

\bibitem{Kasim2022}
\bibinfo{author}{Kasim, M.~F.} \emph{et~al.}
\newblock \bibinfo{journal}{\bibinfo{title}{Building high accuracy emulators
  for scientific simulations with deep neural architecture search}}.
\newblock {\emph{\JournalTitle{Machine Learning: Science and Technology}}}
  \textbf{\bibinfo{volume}{3}}, \bibinfo{pages}{015013},
  \doiprefix\url{10.1088/2632-2153/ac3ffa} (\bibinfo{year}{2021}).

\bibitem{Pfau2020}
\bibinfo{author}{Pfau, D.}, \bibinfo{author}{Spencer, J.~S.},
  \bibinfo{author}{Matthews, A. G. D.~G.} \& \bibinfo{author}{Foulkes, W.
  M.~C.}
\newblock \bibinfo{journal}{\bibinfo{title}{{Ab initio solution of the
  many-electron Schrödinger equation with deep neural networks}}}.
\newblock {\emph{\JournalTitle{Physical Review Research}}}
  \textbf{\bibinfo{volume}{2}}, \bibinfo{pages}{033429},
  \doiprefix\url{10.1103/physrevresearch.2.033429} (\bibinfo{year}{2020}).
\newblock \eprint{1909.02487}.

\bibitem{Kasim2021}
\bibinfo{author}{Kasim, M.~F.} \& \bibinfo{author}{Vinko, S.~M.}
\newblock \bibinfo{journal}{\bibinfo{title}{{Learning the Exchange-Correlation
  Functional from Nature with Fully Differentiable Density Functional
  Theory}}}.
\newblock {\emph{\JournalTitle{Physical Review Letters}}}
  \textbf{\bibinfo{volume}{127}}, \bibinfo{pages}{126403},
  \doiprefix\url{10.1103/physrevlett.127.126403} (\bibinfo{year}{2021}).

\bibitem{Sanchez-Gonzales2021}
\bibinfo{author}{Sanchez-Gonzales, A.} \emph{et~al.}
\newblock \bibinfo{journal}{\bibinfo{title}{{LEARNING GENERAL-PURPOSE CNN-BASED
  SIMULATORS FOR ASTROPHYSICAL TURBULENCE}}}.
\newblock {\emph{\JournalTitle{ICLR 2021 SimDL Workshop}}}
  (\bibinfo{year}{2021}).

\bibitem{wang2020}
\bibinfo{author}{Wang, R.}, \bibinfo{author}{Kashinath, K.},
  \bibinfo{author}{Mustafa, M.}, \bibinfo{author}{Albert, A.} \&
  \bibinfo{author}{Yu, R.}
\newblock \bibinfo{journal}{\bibinfo{title}{Towards physics-informed deep
  learning for turbulent flow prediction}}.
\newblock {\emph{\JournalTitle{arXiv:1911.08655}}}  (\bibinfo{year}{2020}).
\newblock \eprint{1911.08655}.

\bibitem{Zongyi2020}
\bibinfo{author}{Hua, M.}, \bibinfo{author}{Wu, Q.}, \bibinfo{author}{Ng, D.
  W.~K.}, \bibinfo{author}{Zhao, J.} \& \bibinfo{author}{Yang, L.}
\newblock \bibinfo{journal}{\bibinfo{title}{Intelligent reflecting
  surface-aided joint processing coordinated multipoint transmission}}.
\newblock {\emph{\JournalTitle{CoRR}}}
  \textbf{\bibinfo{volume}{abs/2003.13909}} (\bibinfo{year}{2020}).
\newblock \eprint{2003.13909}.

\bibitem{Rackauckas2020}
\bibinfo{author}{Rackauckas, C.} \emph{et~al.}
\newblock \bibinfo{journal}{\bibinfo{title}{{Universal Differential Equations
  for Scientific Machine Learning}}}.
\newblock {\emph{\JournalTitle{arXiv:2001.04385}}}  (\bibinfo{year}{2020}).
\newblock \eprint{2001.04385}.

\bibitem{Kim_2019}
\bibinfo{author}{Kim, B.} \emph{et~al.}
\newblock \bibinfo{journal}{\bibinfo{title}{Deep fluids: A generative network
  for parameterized fluid simulations}}.
\newblock {\emph{\JournalTitle{Computer Graphics Forum}}}
  \textbf{\bibinfo{volume}{38}}, \bibinfo{pages}{59–70},
  \doiprefix\url{10.1111/cgf.13619} (\bibinfo{year}{2019}).

\bibitem{Lusch2018}
\bibinfo{author}{Lusch, B.}, \bibinfo{author}{Kutz, J.} \&
  \bibinfo{author}{Brunton, S.}
\newblock \bibinfo{journal}{\bibinfo{title}{Deep learning for universal linear
  embeddings of nonlinear dynamics}}.
\newblock {\emph{\JournalTitle{Nature Communications}}}
  \textbf{\bibinfo{volume}{9}}, \doiprefix\url{10.1038/s41467-018-07210-0}
  (\bibinfo{year}{2018}).

\bibitem{Sanchez-Gonzalez2018}
\bibinfo{author}{Sanchez-Gonzalez, A.} \emph{et~al.}
\newblock \bibinfo{journal}{\bibinfo{title}{{Graph networks as learnable
  physics engines for inference and control}}}.
\newblock {\emph{\JournalTitle{arXiv:1806.01242}}}  (\bibinfo{year}{2018}).
\newblock \eprint{1806.01242}.

\bibitem{Kochkov2021}
\bibinfo{author}{Kochkov, D.} \emph{et~al.}
\newblock \bibinfo{journal}{\bibinfo{title}{{Machine learning–accelerated
  computational fluid dynamics}}}.
\newblock {\emph{\JournalTitle{Proceedings of the National Academy of
  Sciences}}} \textbf{\bibinfo{volume}{118}}, \bibinfo{pages}{e2101784118},
  \doiprefix\url{10.1073/pnas.2101784118} (\bibinfo{year}{2021}).

\bibitem{Novati2021}
\bibinfo{author}{Novati, G.}, \bibinfo{author}{Laroussilhe, H. L.~d.} \&
  \bibinfo{author}{Koumoutsakos, P.}
\newblock \bibinfo{journal}{\bibinfo{title}{{Automating turbulence modelling by
  multi-agent reinforcement learning}}}.
\newblock {\emph{\JournalTitle{Nature Machine Intelligence}}}
  \textbf{\bibinfo{volume}{3}}, \bibinfo{pages}{87--96},
  \doiprefix\url{10.1038/s42256-020-00272-0} (\bibinfo{year}{2021}).

\bibitem{pathak2020}
\bibinfo{author}{Pathak, J.} \emph{et~al.}
\newblock \bibinfo{journal}{\bibinfo{title}{Using machine learning to augment
  coarse-grid computational fluid dynamics simulations}}.
\newblock {\emph{\JournalTitle{arXiv:2010.00072}}}  (\bibinfo{year}{2020}).
\newblock \eprint{2010.00072}.

\bibitem{SIRIGNANO2020}
\bibinfo{author}{Sirignano, J.}, \bibinfo{author}{MacArt, J.~F.} \&
  \bibinfo{author}{Freund, J.~B.}
\newblock \bibinfo{journal}{\bibinfo{title}{Dpm: A deep learning pde
  augmentation method with application to large-eddy simulation}}.
\newblock {\emph{\JournalTitle{Journal of Computational Physics}}}
  \textbf{\bibinfo{volume}{423}}, \bibinfo{pages}{109811},
  \doiprefix\url{https://doi.org/10.1016/j.jcp.2020.109811}
  (\bibinfo{year}{2020}).

\bibitem{Um2020}
\bibinfo{author}{Um, K.}, \bibinfo{author}{Brand, R.}, \bibinfo{author}{Fei,
  Y.~R.}, \bibinfo{author}{Holl, P.} \& \bibinfo{author}{Thuerey, N.}
\newblock \bibinfo{title}{Solver-in-the-loop: Learning from differentiable
  physics to interact with iterative pde-solvers}.
\newblock In \bibinfo{editor}{Larochelle, H.}, \bibinfo{editor}{Ranzato, M.},
  \bibinfo{editor}{Hadsell, R.}, \bibinfo{editor}{Balcan, M.~F.} \&
  \bibinfo{editor}{Lin, H.} (eds.) \emph{\bibinfo{booktitle}{Advances in Neural
  Information Processing Systems}}, vol.~\bibinfo{volume}{33},
  \bibinfo{pages}{6111--6122} (\bibinfo{publisher}{Curran Associates, Inc.},
  \bibinfo{year}{2020}).

\bibitem{Xie2019}
\bibinfo{author}{Xie, C.}, \bibinfo{author}{Wang, J.}, \bibinfo{author}{Li,
  H.}, \bibinfo{author}{Wan, M.} \& \bibinfo{author}{Chen, S.}
\newblock \bibinfo{journal}{\bibinfo{title}{Artificial neural network mixed
  model for large eddy simulation of compressible isotropic turbulence}}.
\newblock {\emph{\JournalTitle{Physics of Fluids}}}
  \textbf{\bibinfo{volume}{31}}, \doiprefix\url{10.1063/1.5110788}
  (\bibinfo{year}{2019}).

\bibitem{Mueller2020}
\bibinfo{author}{Mueller, T.}, \bibinfo{author}{Hernandez, A.} \&
  \bibinfo{author}{Wang, C.}
\newblock \bibinfo{journal}{\bibinfo{title}{{Machine learning for interatomic
  potential models}}}.
\newblock {\emph{\JournalTitle{The Journal of Chemical Physics}}}
  \textbf{\bibinfo{volume}{152}}, \bibinfo{pages}{050902},
  \doiprefix\url{10.1063/1.5126336} (\bibinfo{year}{2020}).

\bibitem{Roekeghem2016}
\bibinfo{author}{Roekeghem, A.~v.}, \bibinfo{author}{Carrete, J.},
  \bibinfo{author}{Oses, C.}, \bibinfo{author}{Curtarolo, S.} \&
  \bibinfo{author}{Mingo, N.}
\newblock \bibinfo{journal}{\bibinfo{title}{{High-Throughput Computation of
  Thermal Conductivity of High-Temperature Solid Phases: The Case of Oxide and
  Fluoride Perovskites}}}.
\newblock {\emph{\JournalTitle{Physical Review X}}}
  \textbf{\bibinfo{volume}{6}}, \bibinfo{pages}{041061},
  \doiprefix\url{10.1103/physrevx.6.041061} (\bibinfo{year}{2016}).
\newblock \eprint{1606.03279}.

\bibitem{Juneja2019}
\bibinfo{author}{Juneja, R.}, \bibinfo{author}{Yumnam, G.},
  \bibinfo{author}{Satsangi, S.} \& \bibinfo{author}{Singh, A.~K.}
\newblock \bibinfo{journal}{\bibinfo{title}{{Coupling the High-Throughput
  Property Map to Machine Learning for Predicting Lattice Thermal
  Conductivity}}}.
\newblock {\emph{\JournalTitle{Chemistry of Materials}}}
  \textbf{\bibinfo{volume}{31}}, \bibinfo{pages}{5145--5151},
  \doiprefix\url{10.1021/acs.chemmater.9b01046} (\bibinfo{year}{2019}).

\bibitem{Cranmer2020}
\bibinfo{author}{Cranmer, M.} \emph{et~al.}
\newblock \bibinfo{journal}{\bibinfo{title}{{Discovering Symbolic Models from
  Deep Learning with Inductive Biases}}}.
\newblock {\emph{\JournalTitle{arXiv:2006.11287}}}  (\bibinfo{year}{2020}).
\newblock \eprint{2006.11287}.

\bibitem{Udrescu2020}
\bibinfo{author}{Udrescu, S.-M.} \& \bibinfo{author}{Tegmark, M.}
\newblock \bibinfo{journal}{\bibinfo{title}{{AI Feynman: A physics-inspired
  method for symbolic regression}}}.
\newblock {\emph{\JournalTitle{Science Advances}}}
  \textbf{\bibinfo{volume}{6}}, \bibinfo{pages}{eaay2631},
  \doiprefix\url{10.1126/sciadv.aay2631} (\bibinfo{year}{2020}).

\bibitem{godunov59}
\bibinfo{author}{Godunov, S.~K.}
\newblock \bibinfo{journal}{\bibinfo{title}{{A Difference Scheme for Numerical
  Solution of Discontinuous Solution of Hydrodynamic Equations}}}.
\newblock {\emph{\JournalTitle{Mat. Sbornik.}}} \textbf{\bibinfo{volume}{47}},
  \bibinfo{pages}{271--306} (\bibinfo{year}{1959}).

\bibitem{Tikhonov1963}
\bibinfo{author}{Tikhonov, A.~N.}
\newblock \bibinfo{title}{On the regularization of ill-posed problems}.
\newblock In \emph{\bibinfo{booktitle}{Doklady Akademii Nauk}}, vol.
  \bibinfo{volume}{153}, \bibinfo{pages}{49--52}
  (\bibinfo{organization}{Russian Academy of Sciences}, \bibinfo{year}{1963}).

\bibitem{Cullum1971}
\bibinfo{author}{Cullum, J.}
\newblock \bibinfo{journal}{\bibinfo{title}{Numerical differentiation and
  regularization}}.
\newblock {\emph{\JournalTitle{SIAM Journal on Numerical Analysis}}}
  \textbf{\bibinfo{volume}{8}}, \bibinfo{pages}{254--265},
  \doiprefix\url{10.1137/0708026} (\bibinfo{year}{1971}).
\newblock \eprint{https://doi.org/10.1137/0708026}.

\bibitem{Eilers2003}
\bibinfo{author}{Eilers, P.~H.}
\newblock \bibinfo{journal}{\bibinfo{title}{A perfect smoother}}.
\newblock {\emph{\JournalTitle{Analytical chemistry}}}
  \textbf{\bibinfo{volume}{75}}, \bibinfo{pages}{3631--3636}
  (\bibinfo{year}{2003}).

\bibitem{Chartrand2005}
\bibinfo{author}{Chartrand, R.}
\newblock \bibinfo{journal}{\bibinfo{title}{Numerical differentiation of noisy,
  nonsmooth data}}.
\newblock {\emph{\JournalTitle{ISRN Applied Mathematics}}}
  \textbf{\bibinfo{volume}{2011}}, \doiprefix\url{10.5402/2011/164564}
  (\bibinfo{year}{2011}).

\bibitem{Knowles2014}
\bibinfo{author}{Knowles, I.} \& \bibinfo{author}{Renka, R.~J.}
\newblock \bibinfo{journal}{\bibinfo{title}{Methods for numerical
  differentiation of noisy data}}.
\newblock {\emph{\JournalTitle{Electronic Journal of Differential Equations}}}
  \textbf{\bibinfo{volume}{21}}, \bibinfo{pages}{235--246}
  (\bibinfo{year}{2014}).

\bibitem{Battaglia2018}
\bibinfo{author}{Battaglia, P.~W.} \emph{et~al.}
\newblock \bibinfo{journal}{\bibinfo{title}{{Relational inductive biases, deep
  learning, and graph networks}}}.
\newblock {\emph{\JournalTitle{arXiv:1806.01261}}}  (\bibinfo{year}{2018}).
\newblock \eprint{1806.01261}.

\bibitem{Wu2021}
\bibinfo{author}{Wu, Z.} \emph{et~al.}
\newblock \bibinfo{journal}{\bibinfo{title}{{A Comprehensive Survey on Graph
  Neural Networks}}}.
\newblock {\emph{\JournalTitle{IEEE Transactions on Neural Networks and
  Learning Systems}}} \textbf{\bibinfo{volume}{32}}, \bibinfo{pages}{4--24},
  \doiprefix\url{10.1109/tnnls.2020.2978386} (\bibinfo{year}{2021}).
\newblock \eprint{1901.00596}.

\bibitem{vanLeer1977}
\bibinfo{author}{Van~Leer, B.}
\newblock \bibinfo{journal}{\bibinfo{title}{Towards the ultimate conservative
  difference scheme. iv. a new approach to numerical convection}}.
\newblock {\emph{\JournalTitle{Journal of computational physics}}}
  \textbf{\bibinfo{volume}{23}}, \bibinfo{pages}{276--299}
  (\bibinfo{year}{1977}).

\bibitem{Petersen2021}
\bibinfo{author}{Petersen, B.~K.} \emph{et~al.}
\newblock \bibinfo{title}{Deep symbolic regression: Recovering mathematical
  expressions from data via risk-seeking policy gradients}.
\newblock In \emph{\bibinfo{booktitle}{Proc. of the International Conference on
  Learning Representations}} (\bibinfo{year}{2021}).

\bibitem{RFF}
\bibinfo{author}{Rahimi, A.} \& \bibinfo{author}{Recht, B.~H.}
\newblock \bibinfo{journal}{\bibinfo{title}{{Random Features for Large-Scale
  Kernel Machines}}}.
\newblock {\emph{\JournalTitle{NIPS'07: Proceedings of the 20th International
  Conference on Neural Information Processing Systems}}}
  \bibinfo{pages}{1177–1184} (\bibinfo{year}{2007}).

\bibitem{Tancik2020}
\bibinfo{author}{Tancik, M.} \emph{et~al.}
\newblock \bibinfo{journal}{\bibinfo{title}{{Fourier Features Let Networks
  Learn High Frequency Functions in Low Dimensional Domains}}}.
\newblock {\emph{\JournalTitle{arXiv:2006.10739}}}  (\bibinfo{year}{2020}).
\newblock \eprint{2006.10739}.

\bibitem{liaw2018tune}
\bibinfo{author}{Liaw, R.} \emph{et~al.}
\newblock \bibinfo{journal}{\bibinfo{title}{Tune: A research platform for
  distributed model selection and training}}.
\newblock {\emph{\JournalTitle{arXiv preprint arXiv:1807.05118}}}
  (\bibinfo{year}{2018}).

\bibitem{jax2018github}
\bibinfo{author}{Bradbury, J.} \emph{et~al.}
\newblock \bibinfo{title}{{\it JAX}: composable transformations of
  {P}ython+{N}um{P}y programs; http://github.com/google/jax}
  (\bibinfo{year}{2018}).

\bibitem{haiku2020github}
\bibinfo{author}{Hennigan, T.}, \bibinfo{author}{Cai, T.},
  \bibinfo{author}{Norman, T.} \& \bibinfo{author}{Babuschkin, I.}
\newblock \bibinfo{title}{{\it Haiku}: {S}onnet for {JAX};
  http://github.com/deepmind/dm-haiku} (\bibinfo{year}{2020}).

\bibitem{optax2021github}
\bibinfo{author}{Hessel, M.} \emph{et~al.}
\newblock \bibinfo{title}{{\it Optax}: composable gradient transformation and
  optimisation, in jax; http://github.com/deepmind/optax}
  (\bibinfo{year}{2020}).

\bibitem{Kingma2015}
\bibinfo{author}{Kingma, D.~P.} \& \bibinfo{author}{Lei~Ba, J.}
\newblock \bibinfo{journal}{\bibinfo{title}{{ADAM: A METHOD FOR STOCHASTIC
  OPTIMIZATION}}}.
\newblock {\emph{\JournalTitle{arXiv:1412.6980v9}}}  (\bibinfo{year}{2015}).
\newblock \eprint{1412.6980v9}.

\end{thebibliography}

\section*{Acknowledgements}
The authors would like to thank A.R. Bell for comments on the manuscript,
M. Kasim for discussion about ML, and 
B.K. Petersen for their publicly available DSO package.
The authors would also like to acknowledge the use of the University of Oxford Advanced Research Computing (ARC) facility in carrying out this work. http://dx.doi.org/10.5281/zenodo.22558.

\section*{Author contributions statement}
FM developed the methodology, carried out all the research and drafted the 
first version of the manuscript which both authors have reviewed and
contributed to, while GG provided the initial idea of using a graph-network 
to study heat transport.

\section*{Additional information}
\textbf{Competing interests} \\
The authors declare no competing interests.

    \onecolumn

    \begin{table}
    \centering
    \begin{tabular}{ccccc}
    \hline 
    \hline 
     & Parameter & Min & Max & Spacing \\
    \hline
    \multicolumn{1}{l}{Full Grid} \\
    & $n_e$ [$\rm cm^{-3}$]   & $10^{-5}$     & $10^{-3}$     & uniform \\
    & $T_e$ [keV]             & 1             & 10            & uniform \\
    & $\beta_e$               & $10^{-1}$     & $10$     & uniform \\
    & $\epsilon\, L_T/\lambda_{ei}$               & $7.0 \times 10^{-4}$     & $1.0$     & - \\
    \cline{1-5}
    \multicolumn{1}{l}{Gradient Grid PR-20} \\ 
    & $n_e$ [$\rm cm^{-3}$]   & $5.6 \times 10^{-5}$     & $9.5 \times 10^{-4}$     & uniform \\
    & $T_e$ [keV]             & 1.8         & 9.2            & uniform \\
    & $\beta_e$               & $5.6 \times 10^{-1}$     & $9.5$     & uniform \\
    & $\epsilon\, L_T/\lambda_{ei}$               & $4.8 \times 10^{-3}$     & $0.9$     & - \\
    \cline{1-5}
    \multicolumn{1}{l}{Gradient Grid PR-10} \\
    & $n_e$ [$\rm cm^{-3}$]   & $9.6\times 10^{-5}$     & $9.1 \times 10^{-4}$     & uniform \\
    & $T_e$ [keV]             & 2.4             & 8.6            & uniform \\
    & $\beta_e$               & $9.6 \times 10^{-1}$     & $9.1$     & uniform \\
    & $\epsilon\, L_T/\lambda_{ei}$               & $9.6 \times 10^{-3}$     & $0.8$     & - \\
    \cline{1-5}
    \multicolumn{1}{l}{Gradient Grid PR-5} \\
     & $n_e$ [$\rm cm^{-3}$]   & $1.6 \times 10^{-4}$     & $8.5 \times 10^{-4}$     & uniform \\
    & $T_e$ [keV]             & 3.3             & 7.8            & uniform \\
    & $\beta_e$               & $1.6$     & $8.5$     & uniform \\
    & $\epsilon\, L_T/\lambda_{ei}$               & $0.02$     & $0.6$     & - \\
    \hline
    \end{tabular}
    \caption{Grid of parameter space:
    range and spacing for the grids of plasma parameter 
    values at which the heat flux function (top) and its gradient (bottom three panels)
    are evaluated at different sampling density ($N_{ppd}$).
    For each Table section, the last line shows the range of values of the
    heat-flux suppression factor.
    The temperature gradient length is fixed at $L_T=3\times 10^{22}$ cm.
    \label{tab:plasmaparams}}
    \end{table} 
    \begin{table}
    \centering
    \begin{tabular}{ccccccccc}
    \toprule
    \hline 
    \hline 
    Name & $\sigma_n$ & $N_{\rm ppd}$ & $N_{\rm buf}$ & $N_q$ &  \multicolumn{4}{c}{$N_{\nabla q}$} \\
    & ($\times 100$) & & & & Total & Training  & Eval.  & Test \\ [1ex]
    \hline 
     A.0     &  0   & 20 & 4 &  46464  & 32000 &  21760 &  5440 &  4800 \\
     A.1     &  1   & 20 & 4 &  46464  & 32000 &  21760 &  5440 &  4800 \\
     A.5     &  5   & 20 & 4 &  46464  & 32000 &  21760 &  5440 &  4800 \\
     A.10    &  10  & 20 & 4 &  46464  & 32000 &  21760 &  5440 &  4800 \\
     A.20    &  20  & 20 & 4 &  46464  & 32000 &  21760 &  5440 &  4800 \\ [1ex]
    \hline
     B.0     &   0   & 10 & 4 &  8064 &   4000 &   2720 &   680 &   600 \\
     B.1     &   1   & 10 & 4 &  8064 &   4000 &   2720 &   680 &   600 \\
     B.5     &   5   & 10 & 4 &  8064 &   4000 &   2720 &   680 &   600 \\
     B.10    &   10  & 10 & 4 &  8064 &   4000 &   2720 &   680 &   600 \\
     B.20    &   20  & 10 & 4 &  8064 &   4000 &   2720 &   680 &   600 \\ [1ex] 
    \hline
     C.1     & 1  & 5 & 4 &  1764  &  500 &   340 &   85 &  75 \\
     C.0     & 0  & 5 & 4 &  1764  &  500 &   340 &   85 &  75 \\
     C.5     & 5  & 5 & 4 &  1764  &  500 &   340 &   85 &  75 \\
     C.10    & 10 & 5 & 4 &  1764  &  500 &   340 &   85 &  75 \\
     C.20    & 20 & 5 & 4 &  1764  &  500 &   340 &   85 &  75 \\ [0.5ex] 
    \hline
    \end{tabular}
    \caption{Datasets: the columns represent the datasets' name, 
    the percentage of random relative noise added to the 
    flux function, the sampling density represented by the points-per-decade parameter, 
    the total number of buffer grid points, and the total number of flux function evaluations.
    The last four columns relate to the gradient function datasets (three components each),
    including the total number of evaluations and its partitions into
    training (68\%), evaluation (17) and test (15\%) sets, respectively.
    \label{tab:data}}
    \end{table}
     \begin{table}
     \centering
     \begin{tabular}{cccccccccc}
     \hline
     \hline
     \multicolumn{5}{l}{{\bf A-Series}:  $N_{\rm ppd}$ = 20} \\ [0.5ex] 
     \hline
     \multicolumn{3}{c}{} & \multicolumn{3}{c}{Neural Network} & \multicolumn{2}{c}{Regularization} & \multicolumn{2}{c}{Result}\\ [1ex] 
    Model & Dataset & Learning Rate &\hspace{5pt} Layers & Units & $\sigma_{\rm RFF}$ \hspace{5pt} & Type & Param. &\hspace{5pt} RMS Error & Max Error\\ [1ex] 
    \hline
     MA.0  & A.0  & $1.23\times10^{-3}$ & 4 & 1024    & 0.963 & L1      & $1.02\times 10^{-6}$ & $7.6\times10^{-3}$ & $3.9\times10^{-2}$ \\
     MA.1  & A.1  & $1.80\times10^{-3}$ & 4 & 512     & 0.733 & --    & --                    & $7.5\times10^{-3}$ & $3.4\times10^{-2}$ \\
     MA.5  & A.5  & $3.42\times10^{-3}$ & 5 & 256     & 0.601 & L1      & $6.45\times 10^{-6}$  & $1.1\times10^{-2}$ & $9.4\times10^{-2}$ \\
     MA.10 & A.10 & $1.05\times10^{-3}$ & 6 & 512     & 0.663 & --    & --                & $8.8\times10^{-3}$ & $6.5\times10^{-2}$ \\
     MA.20 & A.20 & $2.53\times10^{-3}$ & 6 & 1024    & 1.723 & L1      & $8.44\times 10^{-7}$  & $1.7\times10^{-2}$ & $3.1\times10^{-1}$ \\ [0.5ex] 
     \hline
     \hline
     \multicolumn{5}{l}{{\bf B-Series}:  $N_{\rm ppd}$ = 10} \\ [0.5ex]
     \hline
     \multicolumn{3}{c}{} & \multicolumn{3}{c}{Neural Network} & \multicolumn{2}{c}{Regularization} & \multicolumn{2}{c}{Result}\\ [1ex] 
    Model & Dataset & Learning Rate &\hspace{5pt} Layers & Units & $\sigma_{\rm RFF}$ \hspace{5pt} & Type & Param. &\hspace{5pt} RMS Error & Max Error\\ [1ex] 
    \hline
     MB.0   & B.0  & $1.13\times10^{-3}$ &  4 & 512  & 0.790      & L2  & $6.49\times 10^{-3}$                      &  $8.7\times10^{-3}$ &  $5.6\times10^{-2}$ \\
     MB.1   & B.1  & $2.88\times10^{-3}$ &  6 & 512  & 0.739      & --  & --                                        &  $9.9\times10^{-3}$ &  $6.3\times10^{-2}$ \\
     MB.5   & B.5  & $1.57\times10^{-3}$ &  5 & 512   & 0.564     & L2  & $8.25\times 10^{-4}$                      &    $9.0\times10^{-3}$ &  $8.0\times10^{-2}$ \\
     MB.10  & B.10 & $2.15\times10^{-3}$ &  4 & 512   & 0.965     & L2  & $1.01\times 10^{-4}$                      &    $9.4\times10^{-3}$ &  $8.1\times10^{-2}$ \\
     MB.20  & B.20 & $2.66\times10^{-3}$ &  6 & 256  & 0.844      & L1  & $9.79\times 10^{-6}$                      &    $1.0\times10^{-2}$ &  $1.0\times10^{-1}$ \\ [0.5ex] 
    \hline
    \hline
    \multicolumn{5}{l}{{\bf C-Series}:  $N_{\rm ppd}$ = 5} \\ [0.5ex] 
    \hline
     \multicolumn{3}{c}{} & \multicolumn{3}{c}{Neural Network} & \multicolumn{2}{c}{Regularization} & \multicolumn{2}{c}{Result}\\ [1ex] 
    Model & Dataset & Learning Rate &\hspace{5pt} Layers & Units & $\sigma_{\rm RFF}$ \hspace{5pt} & Type & Param. &\hspace{5pt} RMS Error & Max Error\\ [1ex] 
    \hline
     MC.0    &  C.0  & $1.30\times10^{-3}$  & 5 & 1024   & 0.444    & L1    & $6.18\times 10^{-7}$                  &    $2.4\times10^{-2}$ & $8.2\times10^{-2}$ \\  
     MC.1    &  C.1 & $3.34\times10^{-3}$   & 5 & 128    & 0.515    & --    & --                                    &    $2.4\times10^{-2}$ & $1.3\times10^{-1}$ \\ 
     MC.5    &  C.5 & $1.89\times10^{-3}$   & 6 & 256    & 0.545    & --    & --                                    &    $2.5\times10^{-2}$ & $2.5\times10^{-1}$ \\ 
     MC.10   & C.10 & $2.03\times10^{-3}$   & 5 & 128    & 0.483    & L1    & $6.82\times 10^{-6}$                  &    $2.7\times10^{-2}$ & $1.9\times10^{-1}$ \\ 
     MC.20   & C.20 & $2.51\times10^{-3}$   & 5 & 128    & 0.547    & L2    & $1.00\times 10^{-2}$                  &    $2.7\times10^{-2}$ & $1.8\times10^{-1}$ \\ [0.5ex] 
    \hline
    \end{tabular}
    \caption{Best models selection: from left to right the columns include 
    the model's name, the name of training set as listed in Table~\ref{tab:data}, 
    the number of hidden layers and units, respectively, the $\sigma_{RFF}$ parameter 
    for generation of Random Fourier Features embeddings, and final the RMS and Max statistics 
    for the {\it evaluation} errors. The table is divided into three subtables, one for 
    each density of sampling parameter, $N_{ppd}$, characterising the 
    training data.
    \label{tab:models}}
    \end{table} 

    \begin{table}
    \centering
    \begin{tabular}{c|ccccc}
    \hline 
    \hline 
    \multicolumn{6}{l}{Function Set = $\{+, -,\times, \div, \log, \exp, {\rm const.} \}$, \quad Iterations = 100 } \\ [0.5ex]
    \hline
    \multicolumn{2}{c}{} &
    \multicolumn{2}{c}{Function error} &  \multicolumn{2}{c}{Gradient error}\\
    \multicolumn{1}{c}{Dataset}& \multicolumn{1}{c}{Symbolic Expression for $\epsilon^{-1}$} &
    \multicolumn{1}{c}{RMS}& \multicolumn{1}{c}{Max} &
    \multicolumn{1}{c}{RMS}& \multicolumn{1}{c}{Max} \\
    \hline 
    A.1  & $(0.9992\cdot x_1 + 1.0008\cdot x_2 + 3.9928)^{1.00055}$ & $7.3\times10^{-4}$  & $1.9\times10^{-3}$ & $1.6\times10^{-3}$  & $3.9\times10^{-3}$ \\
    A.5  & $(1.0977 \cdot x_1 + x_2 + 3.8453919)/\log(0.3062\cdot\log(x_3))$ & $7.3\times10^{-2}$ & $1.6\times10^{-1}$ & $1.3\times10^{-1}$ & $2.5\times10^{-1}$ \\
    A.10 & $(x_1 + x_2 + 3.9388) \cdot e^{0.0013\cdot x_2}$ & $8.3\times10^{-3}$ & $1.5\times10^{-2}$ & $1.3\times10^{-2}$ & $1.4\times10^{-1}$ \\
    A.20 & $(0.9855\cdot x_1 + 1.0145\cdot x_2 + 3.8580)^{1.0108}$ & $1.5\times10^{-2}$ & $3.7\times10^{-2}$ & $3.1\times10^{-2}$ & $7.6\times10^{-2}$ \\
    \hline
    \end{tabular}
    \caption{Results from the symbolic regression analysis. From left to right the column indicate
    the name of the dataset from which the training data were sampled, the found symbolic expression
    corresponding for clarity to $\epsilon^{-1}$, the RMS and Max statistics of the relative error for
    the function and its gradient on a random sample of 10$^6$ entries.
    \label{tab:symbolic}}
    \end{table} 
    
    \begin{table}[ht]
    \centering
    \begin{tabular}{cccc}
    \hline 
    \hline 
    Hyperparameter & Search space & Space type   \\[0.5ex] 
    \hline 
    Number hidden layers  & \{4, 5, 6\}         & exhaustive \\[0.5ex] 
    Number hidden units    & \{128, 256, 512, 1024\}  & exhaustive \\[0.5ex] 
    Learning rate        & [$10^{-4}$, \, 2$\times 10^{-3}$] & log-uniform sampling \\[0.5ex] 
    $\sigma_{\rm RFF}$    & [0.1, \, 5.0]  & log-uniform sampling \\
  \hline
    \end{tabular}
    \caption{Reduced space searched for the final tuning of hyperparameters characterising the MLP model.
    \label{tab:hpsearch}}
    \end{table}

    \clearpage

    \clearpage
    \beginsupplement
    \section*{Supplementary materials}

    \newcommand{\RN}[1]{%
    \textup{\uppercase\expandafter{\romannumeral#1}}%
    }
    \begin{figure}[ht]
    \centering
    \includegraphics[width=1\textwidth,]{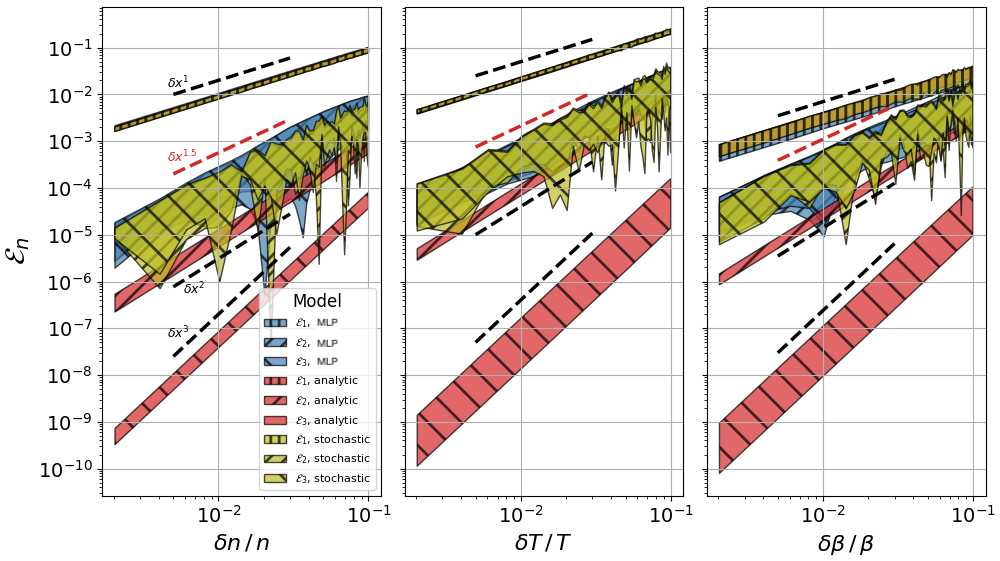}
    \caption{Taylor expansion's error $\RN{2}$:
    same as Fig.~\ref{Fig1} 
    but with superposed data for the function in equation (\ref{eq:stochgrad}), 
    the stochastic model,
    consisting of the analytic flux function $q$ and a random term proportional 
    to its first derivative multiplied by a number randomly sampled within $[-1, ~1]$. 
    \label{FigA1}}
    \end{figure}
    
    \begin{figure}[ht]
    \centering
    \includegraphics[width=1\textwidth,]{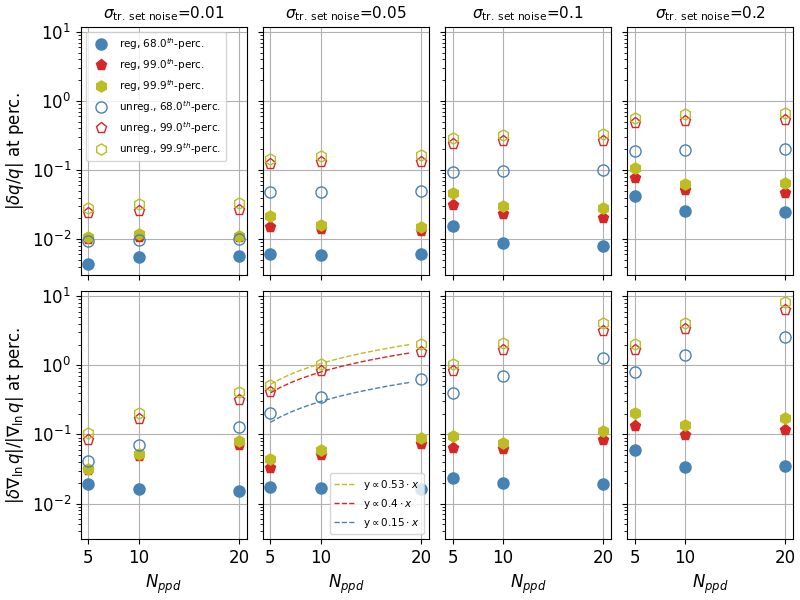}
    \caption{Errors at percentile:
    for each value of the pre-regularization noise, $\sigma_n$,
    the panels show as a function of sampling density represented by 
    the ($N_{ppd}$) parameter, 
    the relative error of the flux function (top) and
    the flux gradient (bottom), the latter estimated
    as the ratio of the Euclidean norm of the flux log-gradient error
    and the Euclidean norm of the correct flux log-gradient (see main text
    for definition of log-gradient and Euclidean norm),
    at 68$^{th}$ (blue circles), 
    99$^{th}$ (red pentagons), and 99.9$^{th}$ (olive hexagons) percentile, 
    respectively. 
    Filled and open symbols refer to regularised and unregularised data, respectively.
    The dashed lines in the second bottom panel from the left is a simple eyeball fit.
    \label{FigA2}
    }
    \end{figure}

    \begin{figure}[ht]
    \centering
    \includegraphics[width=1\textwidth,]{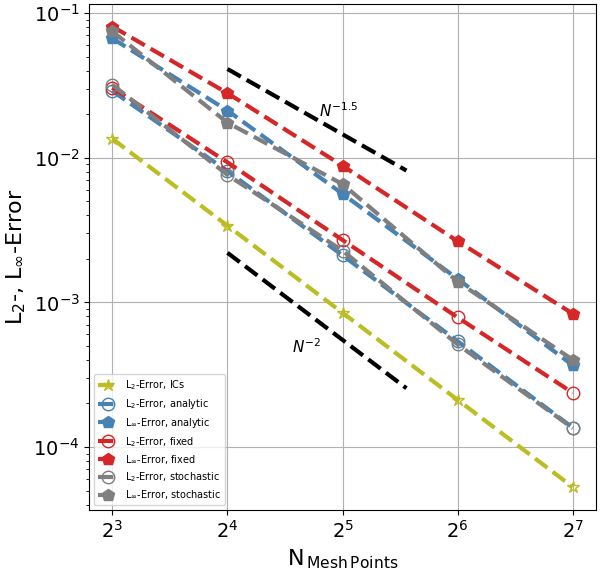}
    \caption{Convergence test $\RN{2}$:
    $L_2$ and $L_\infty$ error norms 
    for the `analytic' (blue), `fixed' (red) and `stochastic' (gray) models.
    The plotted errors are averaged over a sample of 30 runs 
    with different, randomly chosen values of the 
    unperturbed thermodynamic parameters, ($n, T, \beta$).
    The $L_2$ error norm  for the initial conditions (olive) is 
    also shown together with curves representing $N^{-2}$ and $N^{-1.5}$  
    error drop rate.
    \label{FigA3}
    }
    \end{figure}

\end{document}